\documentclass[preprint2]{aastex}

\usepackage{amsmath}

\slugcomment{}

\shorttitle{$E_{p}$-Flux Correlation } \shortauthors{Lu \& Liang}

\begin{document}


\title{The $E_{\rm p}$-Flux Correlation in the Rising and Decaying Phases of Gamma-Ray Burst Pulses: Evidence for Viewing Angle Effect?}


\author{R.-J. Lu\altaffilmark{1}, S.-J. Hou\altaffilmark{1} and En-Wei Liang\altaffilmark{1,2}}
\altaffiltext{1}{Department of Physics, Guangxi University, Guangxi 530004,
China.}\altaffiltext{2}{Department of Physics and Astronomy, University of
Nevada Las Vegas, Las Vegas, NV 89154, USA.}  \email{luruijing@gxu.edu.cn;
lew@gxu.edu.cn}




\begin{abstract}
A time-resolved spectral analysis for a sample of 22 intense, broad GRB pulses
from the {\em CGRO}/BATSE GRB sample is presented. We fit the spectra with the
Band function and investigate the correlation between the observed flux ($F$)
and the peak energy ($E_{\rm p}$) of the $\nu f_\nu$ spectrum in the rising and
decaying phases of these pulses. Two kinds of $E_{\rm p}$ evolution trends,
i.e., hard-to-soft (the two-third pulses in our sample) and $E_{\rm
p}$-tracing-$F$ (the one-third pulses in our sample) are observed in pulses
from different GRBs and even from different pulses of the same burst. No
dependence of spectral evolution feature on the pulse shape is found. A tight
$F-E_{\rm p}$ positive correlation is observed in the decaying phases, with a
power-law index $\sim 2.2$, which is much shallower than that expectation of
the curvature effect. In the rising phase, the observed $F$ is either
correlated or anti-correlated with $E_{\rm p}$, depending on the spectral
evolution feature, and the power-law index of the correlation is dramatically
different among pulses. More than $80\%$ of the low energy photon indices in
the time-resolved spectra whose $E_{\rm p}$ is anti-correlated with $F$ during
the rising phase violate the death line of the synchrotron radiation,
disfavoring the synchrotron radiation model for these gamma-rays. The $F-E_{\rm
p}$ correlation, especially for those GRBs with $E_{\rm p}$-tracking-$F$
spectral evolution, may be due to the viewing angle and jet structure effects.
In this scenario, the observed $F-E_{\rm p}$ correlation in the rising phase
may be due to the line of sight from off-beam to on-beam toward a structured
jet (or jitter),  and the decaying phase is contributed by both the on-beam
emission and the decayed photons from high latitude of the GRB fireball,
resulting in a shallower slope of the observed $F-E_{\rm p}$ correlation than
that predicted by the pure curvature effect.
\end{abstract}


\keywords{gamma-ray burst: general---radiation mechanisms:
non-thermal---methods: data analysis}



\section{Introduction}
The physics of prompt emission of gamma-ray bursts (GRBs) remains as a great
puzzle. Analysis for a large sample of GRB spectra observed with Burst and
Transient Source Experiment (BATSE) on board {\em CGRO} reveals that the GRB
spectra are non-thermal, well-fit with a smoothly-jointed broken power-law, the
so-called Band function (Band et al. 1993). The physical radiation mechanism
that shapes such a spectrum shape is unclear (e.g., Zhang \& M\'{e}sz\'{a}ros
et al. 2004). The peak energy of the $\nu f_\nu$ spectrum ($E_{\rm p}$) is one
of the most interesting parameters of GRBs. The relation between the burst
energy and $E_{\rm p}$ may shed light on the physics of GRBs (Zhang \& M
\'{e}sz\'{a}ros 2002). With a sample of 12 GRBs detected by BeppoSAX, Amati et
al. (2002) found a positive correlation between the isotropic gamma-ray energy
$(E_{\rm iso})$ and the peak energy in the burst frame ($E_{\rm p,z}$). This
correlation was confirmed and even extended to the X-ray flashes by the
observations with {\em HETE}-2 and {\em Swift} (Sakamoto et al., 2006; Amati
2006). By geometrical-correction for the jet beaming effect or empirically
incorporating the breaks in the optical afterglow lightcurves, this correlation
is even getting tighter (Ghirlanda et al. 2004; Liang \& Zhang 2005).
Similarly, the isotropic peak luminosity ($L_p$) is correlated with $E_{p,z}$
(Wei et al. 2003; Yonetoku et al. 2004). The average flux ($F$) in a given
epoch is also correlated with the $E_{\rm p}$ in the corresponding epoch within
a GRB (Liang et al. 2004). The $F-E_{\rm p}$ correlation suggests that the
$E_{\rm p}-E_{\rm iso}$ and the $E_{\rm p}-L_{p}$ correlations would not be due
to the observational selection effects (c.f., Nakar \& Piran 2005; Band \&
Preece 2005; Shahmoradi \& Nemiroff 2009). All the prompt GRB emission models
predict $E_{\rm p}$ as a function of both $E_{{\rm iso}}$ (or $L_{\rm p}$) and
the initial Lorentz factor of the GRB fireball ($\Gamma_0$) (e.g. Table 1 of
Zhang \& M\'{e}sz\'{a}ros 2002). Most recently, Liang et al. (2010) discover a
tight correlation between $E_{\rm iso}$ and $\Gamma_0$, i.e.,
$\Gamma_0=182(E_{\rm iso}/10^{52}\rm \ erg)^{0.25}$. This correlation poses
constraints on prompt emission models. For example, the internal shock
synchrotron model, the most favorite model for GRBs, predicts $E_{\rm p}
\propto L^{1/2} \Gamma_0^{-2}$ (Zhang \& M\'{e}sz\'{a}ros 2002). Combining with
the trivial proportionality of $L\propto E_{\rm iso}$, one can find that
$E_{\rm p}$ should not depend on $L$, indicating that the $E_{\rm p}-E_{\rm
iso}$ correlation may not be explained in the framework of the internal shock
synchrotron model.

This paper dedicates to revisit the $F-E_{\rm p}$ correlation and its possible
physical origin. The $E_{\rm p}$ of a given burst evolves with time, tracing
with the lightcurve ($E_{\rm p}$-tracing-$F$) or evolving as hard-to-soft
(Liang \& Kargatis 1996; Ford et al. 1995; Kaneko et al. 2006). Intuitively,
the $F-E_{\rm p}$ correlation should be the foundation of the Amati relation,
if both are based on the same physical origin. Since those GRBs having
hard-to-soft spectral evolution do not have a coherent $F-E_{\rm p}$
correlation, they should violate the Amati relation. However, the current
sample of GRBs with both $E_{\rm iso}$ and $E_{\rm p}$ measurements well follow
the Amati-relation (Amati et al. 2009), regardless of the spectral evolution
feature of these GRBs. For example, although the spectrum of GRB 060218 evolves
as hard-to-soft (Toma et al. 2007; Dong et al. 2010), it well satisfies the
Amati-relation (Amati et al. 2006). Therefore, the Amati relation seems to be
unrelated to the dependence of $F$ on $E_p$. This gives rise to a big puzzle of
the physical origin of the Amati-relation.

It is well known that the GRB lightcurvs are generally composed of some
overlapped pulses. An individual shock episode may give rise to a pulse, and
the random superposition of many such pulses results in the observed complexity
of GRB light curves. As the building blocks of GRB lightcurves, the broad,
well-separated pulses are good candidates to reveal the physics of GRBs. The
$E_{\rm p}$ evolution within GRB pulses has been extensively studied,
especially for the decaying phase of pulses (Golenetskii et al. 1983; Norris et
al. 1986; Kargatis et al. 1994, 1995; Bhat et al. 1994; Crider et al. 1999;
Peng et al. 2009; Lu \& Liang 2010). While the decaying phase of a GRB pulse
might be contributed by the curvature effect of the GRB fireball (e.g.,
Fenimore et al. 1995; Kobayashi et al. 1997; Qin et al. 2002; Dermer 2004; Shen
et al. 2005), both the temporal and spectral behaviors of the rise phase of a
pulse may depend on the dynamics of the GRB fireball, the electron
acceleration, and the radiation mechanism (e.g., Kobayashi et al. 1997). The
$F-E_{\rm p}$ correlations in the two phases, if any, should dramatically
different, which was shown by Ohno et al. 2009 and Ghirlanda et al. (2010) for
some bright GRBs. We here present a detailed analysis on the $F-E_{\rm p}$
relation in both the rising and decaying phases of smooth, broad GRB pulses
observed with {\em CGRO}/BATSE.

We present our sample and spectral analysis in Section 2. The lightcurves and
the $E_p$ evolution as well as the $F-E_{\rm p}$ correlation are shown in
Section 3. Since the low energy photon index of the Band function is critical
to justify if the radiation is from synchrotron radiation, we present the
distribution of the low energy photon indices for those spectra having an
anti-correlation between $E_{\rm p}$ and $F$ in Section 4. We discuss the
possible physical origin of the $F-E_{\rm p}$ correlation (or anti-correlation
) in Section 5. Conclusions are presented in Section 6.


%
%
%
%
%
%
%
%
%

\section{Sample selection and time-resolved spectral analysis}
We make use the data observed with BATSE\footnote{Although {\em Fermi}/GBM has
established a large sample of GRBs in similar energy band as {\em CGRO}/BATSE,
we only use the data observed with BATSE in this analysis since they are
well-studied and the uncertainty of background subtraction for GBM data make
our time-resolved spectral fits has larger uncertainty than BATSE data
(especially in the rising phases). }. Kaneko et al. (2006) presented a sample
of 8459 time-resolved burst spectra for 350 bright GRBs observed with BATSE.
Our sample of GRB pulses are taken from this sample. We download the spectra
data from the web site http://www.batse.msfc.nasa.gov/~kaneko/. We first select
bright pulses from the lightcurves of these GRBs. Technically, it is difficult
to define a genuine pulse from GRB lightcurves. Flickering and/or superimposing
weak pulses make complication to employ a rigid criterion to select our pulse
sample. We use the same criteria as that described in Liang et al. (2002) to
select our sample. We make time-resolved analysis for these pulses with RMFIT,
a package of spectral analysis routines (version 3.3) developed by the BATSE
team (Mallozzi et al. 2005 and Preece et al. 2008). Although the time-resolved
spectral analysis for these pulses are present in Kaneko et al. (2006), we
re-do the time-resolved spectral analysis for these spectra by lowering down a
little bit of the signal-to-noise ratio (SNR) in each time slice in order to
get more slices in both the rising and decaying phases for our analysis. We
adopt $SNR=30$, compared to $SNR=45$ in Kaneco et al. (2006). We fit the
spectra with the Band function (Band et al. 1993). The reduced $\chi^2$ of our
fits are normally $\sim 0.9-1.1$. Since we focus on the $F-E_{\rm p}$
correlation in both the rising and decaying parts of a pulse, our sample
includes only those pulses that are intense and broad enough to make robust
spectral fit for at least three time slices in the rising and decaying
segments, respectively. We finally get a sample of 22 pulses. The average flux
$F$ in each time slide is then derived from the Band model spectral parameters
in the 30-$10^4$ keV band (as done in Yonetoku et al. 2004 and Liang et al.
2004). Our time-resolved spectral analysis results are available in the online
material of this paper.

\section{Temporal evolution of $E_{\rm p}$ and $F-E_{\rm p}$ correlation}
Figure 1 shows the lightcurve with temporal evolution of the $E_{p}$ and the
$F-E_{\rm p}$ correlation for the pulses in our sample. It is found that the
shape of these pulses are semi-symmetric, slightly different from the FRED
pulses usually seen in the GRB lightcurves. This would be due to our sample
selection effect since we include only those pulses that the rising part is
long and bright enough to make time-resolved spectral fit.

The spectral evolution feature is well classified into two groups, i.e.,
hard-to-soft and $E_{\rm p}$-tracing-$F$. Fifteen out of the 22 belong to the
group of the hard-to-soft evolution, including \# 647, 973, 1883, 2083, 2193,
2387, 3658, 5478, 6397, 6504, 6630, 7293, 7475, 7588, and 7771. The others,
including \#1625, 1733, 1956, 3003, 3765, 5523 and 5601, are of the $E_{\rm
p}$-tracing-flux group. No dependence of the spectral evolution feature on the
pulse shape is found.

The observed fluxes in the decaying phases of these pulses are tightly
correlated with $E_{\rm p}$. We fit the $F-E_{\rm p}$ correlation (or
anti-correlation) in the rising and decaying parts with a power-law model
$F\propto E_{\rm p}^{\kappa_r(\kappa_d)}$, where $\kappa_r(\kappa_d)$ is the
power-law index in the rising (decaying) part. The results are reported in
Table 1. As shown in Figure \ref{kappad}, although no universal $\kappa_{d}$
value is found about pulses, the distribution of $\kappa_d$ clustered at $\sim
2.20$ (see also Borgonovo \& Ryde 2001). We should emphasize that the large
dispersion of $\kappa_d$ is not due to the uncertainty of $\kappa_d$
measurement. As shown in Table 1, even considering the the errors of
$\kappa_d$, the $\kappa_d$ of some pulses confidently deviates $\sim 2$. The
dispersion of $\kappa_d$ would be physical (see discussion below). The observed
flux in the rising phase are either correlated or anti-correlated with $E_{\rm
p}$ in the rising phase, depending on the spectral evolution feature. The
power-law index in the $F-E_{\rm p}$ correlation in the rising phase,
$\kappa_r$, spans from $-4$ to 10 without any correlation with the pulse shape
and $\kappa_d$.

It is also very interesting that the $E_{\rm p}$ evolution in different pulses
of a GRB may be also different. Two well-separated pulses are observed in the
BATSE trigger\# 2038. As shown in Figure \ref{two pulses}, the $E_{\rm p}$
evolves as hard-to-soft during the first pulse but it traces the intensity of
the flux in the second pulse. The $F-E_{\rm p}$ correlations in the decaying
phases of the two pulses are similar. However, they are absolutely different in
the rising phases.

\section{Distributions of low-energy photon indices}
The $F-E_{\rm p}$ anti-correlation observed in GRB pulses having a
hard-to-soft spectral evolution is inconsistent with the expectation
of the most favorite GRB model, the synchrotron internal shock
model. The model predicts that the low-energy photon index should
not exceed -2/3, with the assumption that the optical depth of the
shocked material is less than unity (e.g., Preece et al. 1998). We
show the distribution of the best fit $\alpha$ for those
time-resolved spectra whose $E_{\rm p}$ is anti-correlated with $F$
in Figure \ref{alpha}. It is found that 50 out of the 61 spectra
($\sim 82\%$) have a low energy photon index being larger than -2/3.
This percentage is only $32\%$ for those spectra whose $E_{\rm p}$
is positively correlation with $F$. If taking the values of their 1
$\sigma$ lower limits, we find the two percentages become $\sim
78\%$ and $\sim 19\%$, respectively. Therefore, the $F-E_{\rm p}$
anti-correlation cannot be explained with the synchrotron radiation
model.

\section{Implications}
\subsection{$E_{\rm p}$ evolution confronting with the radiation models}
The $E_{\rm p}$ evolution feature within a pulse is essential to study
radiation models of the prompt emission. As shown above, the most common
spectral evolution feature in GRB pulses (the two-third pulses in our sample)
is the hard-to-soft evolution, in which the peak energy of the spectrum
decreases monotonically over the entire pulse (Norris et al. 1986), and the
secondly one (the one-third of the pulses in our sample) is the $E_{\rm
p}$-tracking-$F$ (Golenetskii et al. 1983). More interestingly, the $E_{\rm p}$
in different pulses of \# 2083 even shows different evolution behavior (see
Figure \ref{two pulses}). Different models may design some kinds of $E_{\rm p}$
evolution. However, it is difficult to accommodate two completely different
evolution trends under one mechanism.

The most favorite GRB model is the synchrotron internal shock model. In this
model, a GRB pulse is produced by the collision of two relativistic shells, in
which the rising phase is related to the dynamics and the physical parameters
of the shocked fireball shell and the decay phase is due to the time decay of
photons from the high latitude of the fireball, the so-called curvature effect
(e.g., Kobayshi et al. 1997). Most recently, Zhang \& Yan (2010, in
preparation) proposed that the internal collision may induce magnetic
reconnection and turbulence to explain the prompt gamma-ray emission. In their
model it is expected that $E_{\rm p}$ is positively related to $\sigma$, the
ratio between Poynting flux and baryonic flux. Since the magnetic energy is
continuously converted to the particle energy during an emission episode, the
hard-to-soft $E_{\rm p}$ evolution is naturally expected from their model.

The radiation models discussed above essentially explain the hard-to-soft
emission with the decrease of the energy of radiating particles over the
emission episode. The $E_{\rm p}$-tracking-$F$ evolution, however, is hard to
interpret with these models, especially for different $E_{\rm p}$ evolution
features observed among pulses of a given GRB as shown in Figure \ref{two
pulses} and the observed $F-E_{\rm p}$ anti-correlation. The $\alpha$ of more
than $80\%$ time-resolved spectra whose $E_{\rm p}$ are anti-correlated with
$F$ violate the death line of the synchrotron radiation model.

\subsection{$F-E_{\rm p}$ correlation confronting with viewing angle effect}
It is most likely that the $E_{\rm p}$ evolution and the $F-E_{\rm p}$
correlation are not related to radiation physics. It was also proposed that the
broad pulses in GRBs may be shaped by GRB jet precession (Portegies Zwart et
al.1999; Reynoso et al.2006; Lei et al.2007; Liu et al. 2010). The waggle of
the GRB jet may result in off-beam and on-beam cycle to produce broad pulses
\footnote{It was suggested that the micro-variability in the GRB pulses are
attributed to the turbulence of blast wave (Zhang \& Yan 2010, in preparation)}
and the $E_{\rm p}$-tracking-$F$ spectral evolution (Liu et al. 2010). The
observed lightcurve for initial off-beam to on-beam may rise rapidly to trigger
our detectors, especially in the case of a sharp-edge, highly structured jet
(Panaitescu et al. 1998; Panaitescu \& Vestrand 2008), shaping a
fast-rise-exponential-decay pulse and hard-to-soft spectral evolution since the
early rising part may be not bright enough to trigger our instruments. This may
explain the observed different spectral evolution trends in the pulses of a
burst as shown in Figure \ref{two pulses}.

A clear anti-correlation between $F$ and $E_{\rm p}$ is observed in the rising
phase of a pulse with hard-to-soft spectral evolution, but the power-law index
of the correlation is dramatically different among pulses, indicating that no
universal relation is observed. This may be reasonable since the emission in
the rising phase may be complicated by the dynamics of the fireball, the
radiation mechanisms, the jet structure and viewing angle, micro physical
parameters, etc. However, one is difficult to expect a clear $F-E_{\rm p}$
correlation from stochastic dynamics of the fireball and particle acceleration
process (e.g., Zhang \& M\'{e}sz\'{a}ro 2002). Therefore, the $F-E_{\rm p}$
correlation observed in a given pulse, especially  for those GRBs with $E_{\rm
p}$-tracking-$F$ spectral evolution, may be due to the viewing angle and jet
structure effects, as we discuss above.

In the viewing angle and jet structure dominated scenario, the
contribution of the high latitude photons from the fireball (the
so-called curvature effect) would be increased in the decaying phase
of a pulse and it should dominate the observed flux when the line of
sight moves out the jet edge. Therefore, the mix of the on-axis and
off-axis contributions may result in a shallower decay slope of the
observed $F-E_{\rm p}$ correlation as theoretical prediction. If the
curvature effect dominates the decaying phase of the pulses, the
expected $F-E_{\rm p}$ relation at late time should be $F\propto
E_{p}^3$ (e.g., Fenimore et al. 1995; Kumar \& Panaitescu 2000;
Dermer et al. 2004). As shown in Fig. \ref{kappad}, the $\kappa_d$
of most GRBs in our sample are indeed shallower than the
prediction\footnote{As shown by Zhang et al. (2009), the curvature
effect involving a non-power-law spectrum in the radiating surface
also modify the slope of the $F-E_{\rm p}$ correlation(see also Qin
2008)}.

The $\alpha$ distribution may shed light on the jet structure and radiation
physics. Medvedev (2006) showed that spectrum of jitter radiation from GRB
shocks containing small-scale magnetic fields and propagating at an angle with
respect to the line of sight may vary considerably. The low-energy photon index
may be significantly larger than the death line of the synchrotron radiation,
i.e., -2/3. The $E_{\rm p}$ evolution and $F-E_{\rm p}$ correlation may be
explained with a combined effect of temporal variation of the viewing angle and
relativistic aberration of an individual thin, instantaneously illuminated
shell.

\subsection{$F-E_{\rm p}$ correlation vs the Amati relation}
The correlation between the observed flux and $E_{\rm p}$ within a GRB pulse is
critical to explore the physics of the observed $L-E_{\rm p}$ and $E_{\rm
iso}-E_{\rm p}$ relations (Amati et al. 2002; Liang et al. 2004). The $L-E_{\rm
p}$ and $E_{\rm iso}-E_{\rm p}$ relations would globally  reflect the $F-E_{\rm
p}$ correlation within a GRB. Essentially, they are time-integrated effect of
the $F-E_{\rm p}$ correlation (Firmani et al. 2009). Although the distribution
of $\kappa_d$ for the pulses in our sample has large dispersion, it normally
peaks at $2$. The emission in the decaying phases of all pulses in a GRB should
dominate the total emission of the burst since the duration of the decaying
phase of a pulse if generally much longer than the rising phase. Therefore, one
may observe an $F-E_{\rm p}$ correlation within a GRB or an $E_{\rm iso}-E_{\rm
p}$ correlation among bursts, with a power-law index $\sim 2$. We illustrate
the $F-E_{\rm p}$ correlation within a GRB with multiple pulse in Figure
\ref{Multiple}. An $F-E_{\rm p}$ correlation with a power-law index $\sim 2$ is
clearly seen. Therefore, the $L_p-E_{\rm p}$ or $E_{\rm iso}-E_{\rm p}$
correlations should be dominated by the $F-E_{\rm p}$ correlation. As discussed
above, the $F-E_{\rm p}$ is difficult to explain with the radiation physics.
The $L_p-E_{\rm p}$ and $E_{\rm iso}-E_{\rm p}$ correlations thus may not be
interpreted with the radiation models. This is consistent with that inferred
from the recent discovery of the tight correlation between the $E_{\rm iso}$
and the initial Lorentz factors of the GRB fireball (Liang et al. 2010).

\section{Conclusions}
With our time-resolved spectral analysis for a sample of 22 intense, broad GRB
pulses we find that the $E_{\rm p}$ evolution feature is well classified into
two groups, i.e., hard-to-soft (two-third pulses in our sample) and $E_{\rm p}$
tracing-$F$ (one-third pulses in our sample). Two kinds of spectral
evolutionary trends are also observed in different pulses of a burst. No
dependence of spectral evolution feature on the pulse shape is observed.

A tight $F-E_{\rm p}$ correlation, $F\propto E^{\kappa_d}$, is observed in the
decaying phases of these pulse. Although the $\kappa_{d}$ ranges in a broad
range, from $0.6$ to $ \sim 4.0$, their distribution normally peaks at $\sim
2$, much shallower than that expectation of the curvature effect. In the rising
phase, the observed $F$ is either correlated or anti-correlated with $E_{\rm
p}$, $F\propto E_{\rm p}^{\kappa_r}$, depending on the spectral evolution
feature. The distribution of $\kappa_r$ spans from $-4$ to 10 without any
correlation with the pulse shape and $\kappa_d$. More than $80\%$ of the low
energy photon indices in the time-resolved spectra whose $E_{\rm p}$ is
anti-correlated with $F$ violate the death line of the synchrotron radiation.

The spectral evolution features and the observed $F-E_{\rm p}$ correlation are
difficult to explain with the radiation models. We propose that the observed
the $F-E_{\rm p}$ correlation observed in a given pulse, especially  for those
GRBs with $E_{\rm p}$-tracking-$F$ spectral evolution, may be due to the
viewing angle and jet structure effects. In this scenario, the observed
$F-E_{\rm p}$ correlation in the rising phase is due to the line of sight from
off-beam to on-beam toward a structured jet. The contribution of the high
latitude photons from the fireball(the so-called curvature effect) would be
increased in the decaying phase of a pulse and it should dominate the observed
flux when the line of sight moves out the jet edge. Therefore, the mixed
contributions from the on-beam and the curvature effect may result in a
shallower decay slope of the observed $F-E_{\rm p}$ correlation as theoretical
prediction.

\acknowledgments We appreciate valuable suggestions from the
referee. We also thank Bing Zhang \& Yi-Ping Qin for helpful
discussion. This work is supported by the National Natural Science
Foundation of China under grants No. 10747001 and 10873002, National
Basic Research Program (''973" Program) of China (Grant
2009CB824800), Chinese Academy of Science (under grant
KJCXZ-YW-T19), Guangxi SHI-BAI-QIAN project (Grant 2007201), Guangxi
Science Foundation (2010GXN SFA013112 and 2010GXNSFC013011), the
program for 100 Young and Middle-aged Disciplinary Leaders in
Guangxi Higher Education Institutions, and the research foundation
of Guangxi University (M30520). EWL is a visiting scholar of UNLV
during the revision of this paper with support from NASA NNX09AT66G,
NNX10AD48G, and NSF AST-0908362.

\clearpage




\clearpage
\begin{figure*}
\includegraphics[scale=0.38]{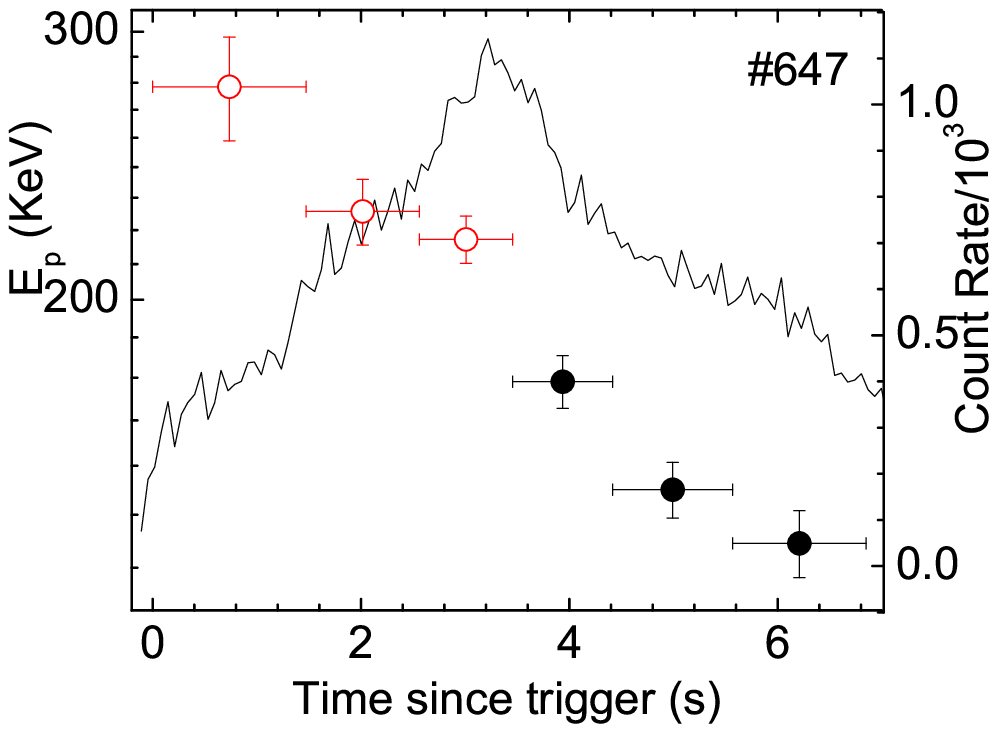}
\includegraphics[scale=0.38]{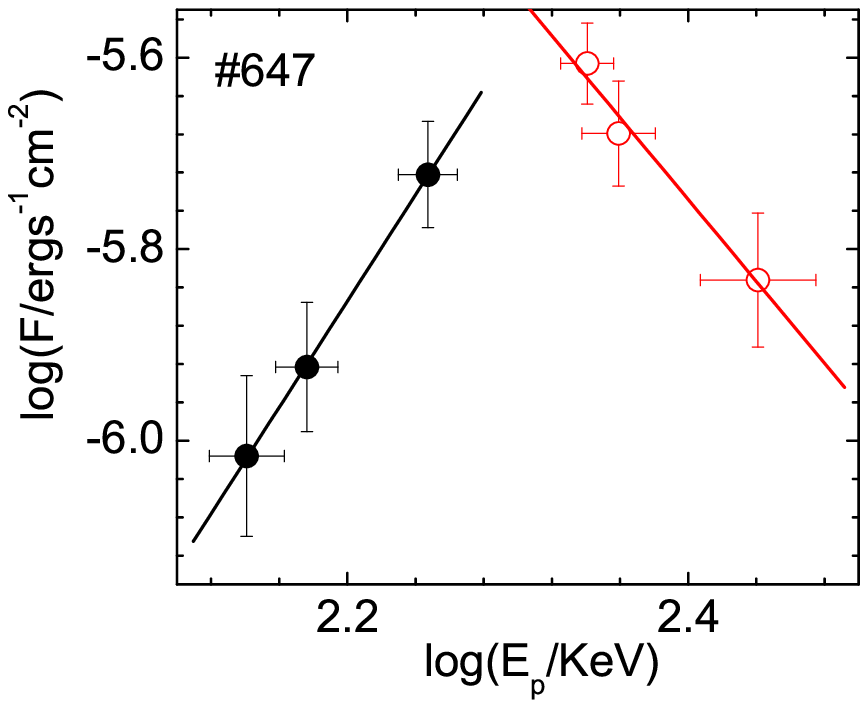}
\includegraphics[scale=0.38]{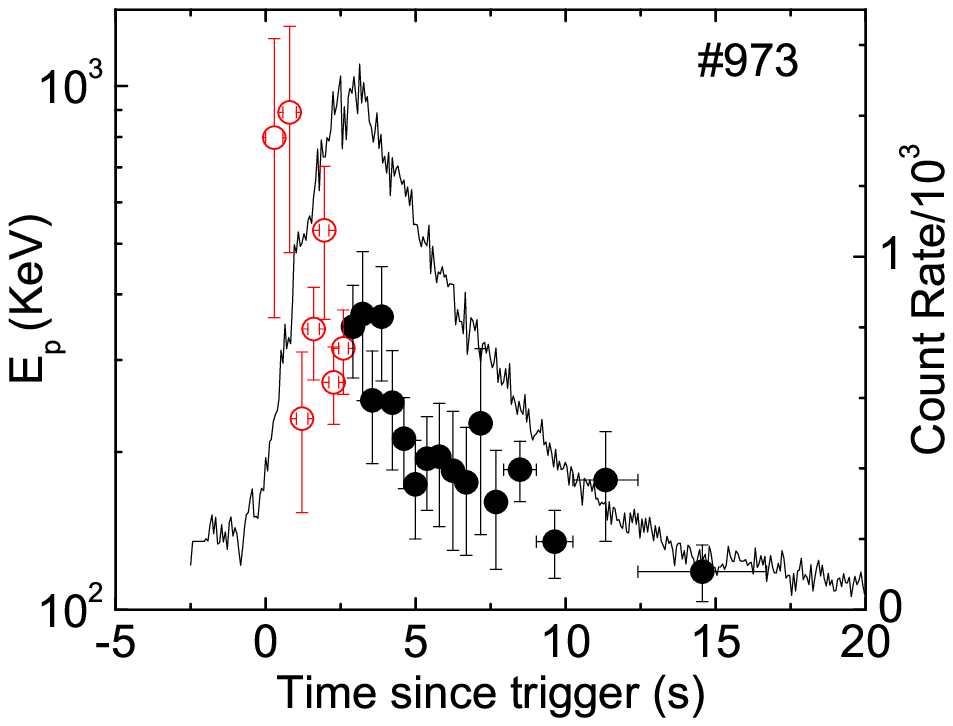}
\includegraphics[scale=0.38]{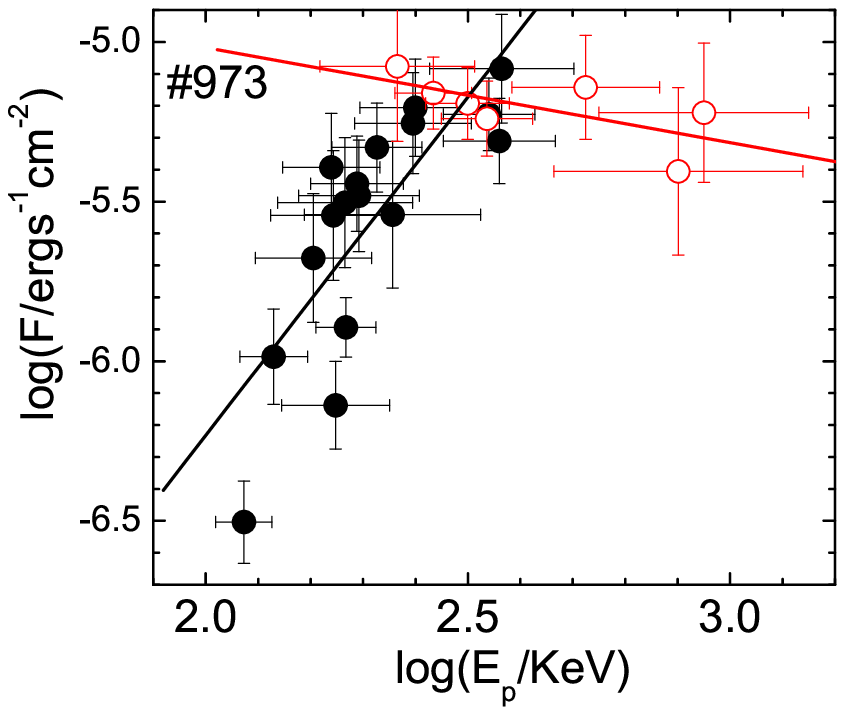}

\includegraphics[scale=0.38]{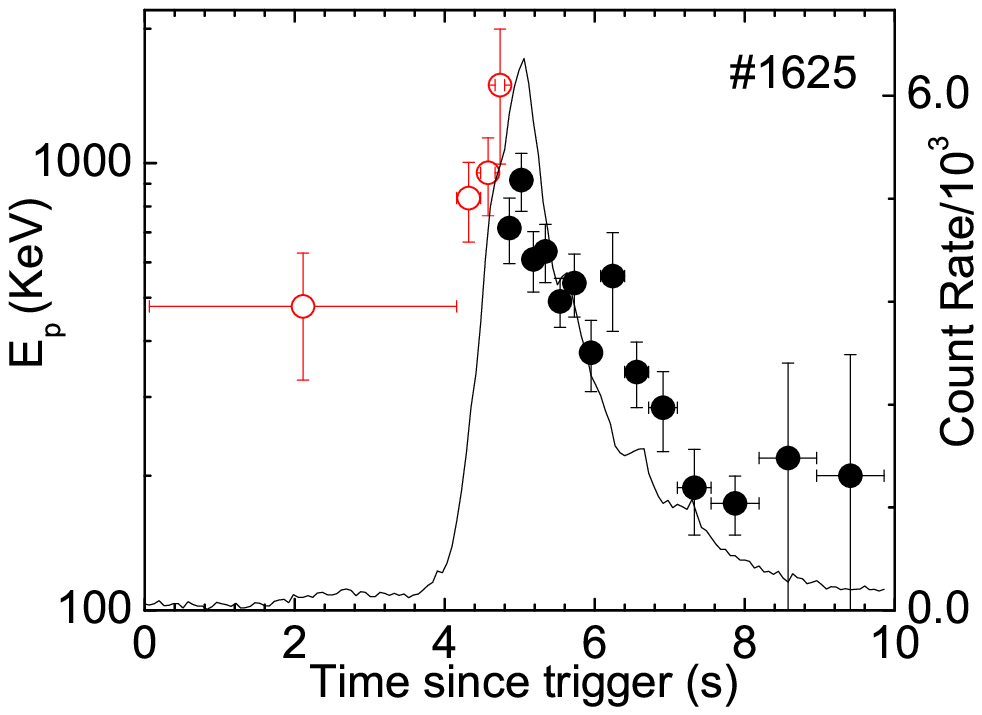}
\includegraphics[scale=0.38]{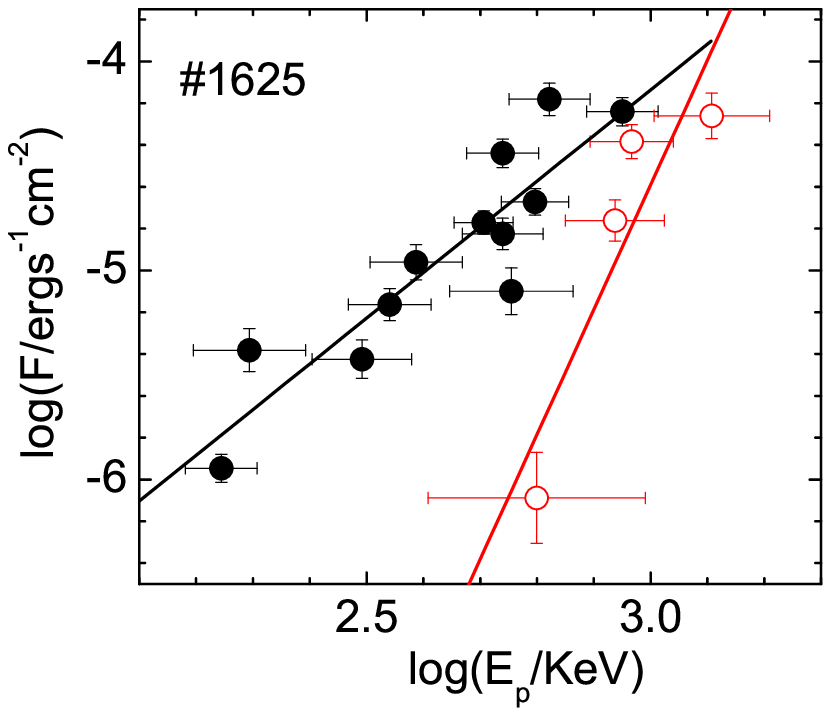}
\includegraphics[scale=0.38]{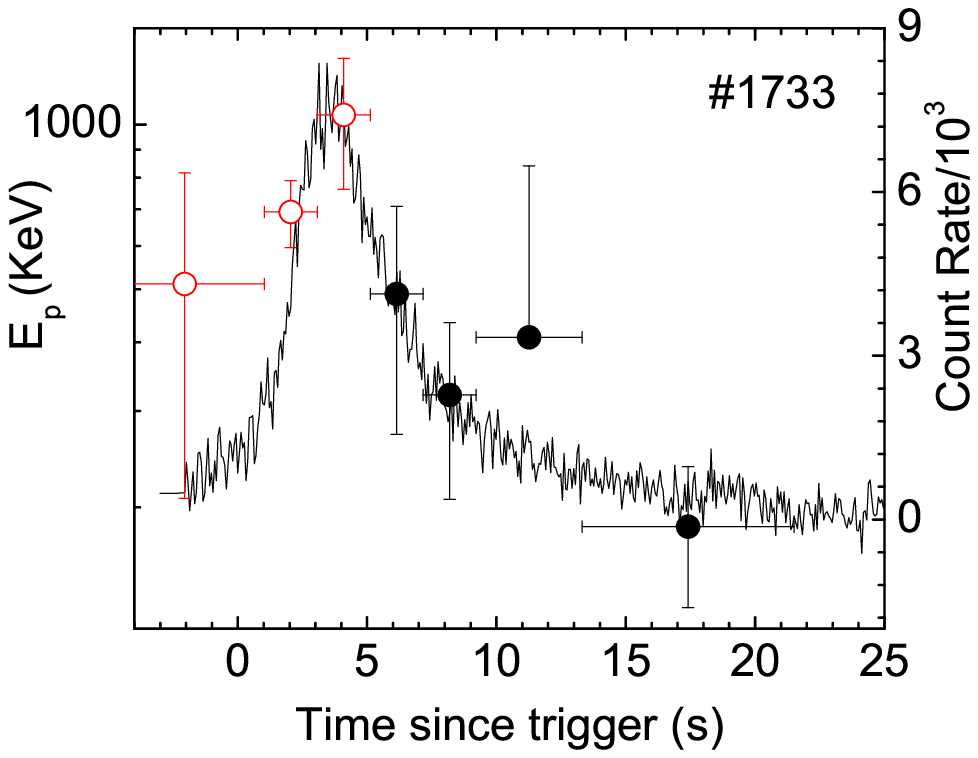}
\includegraphics[scale=0.38]{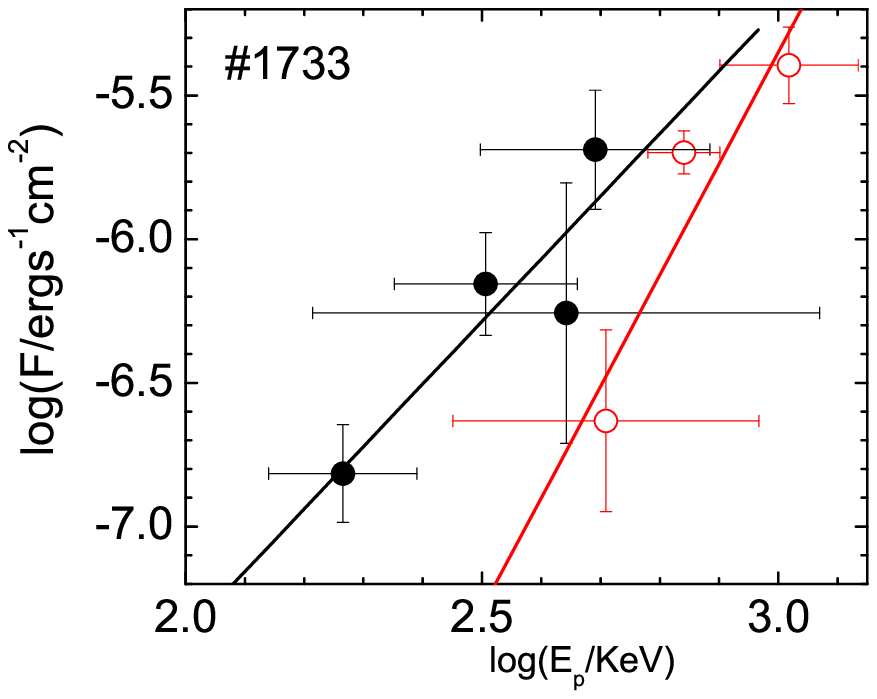}

\includegraphics[scale=0.38]{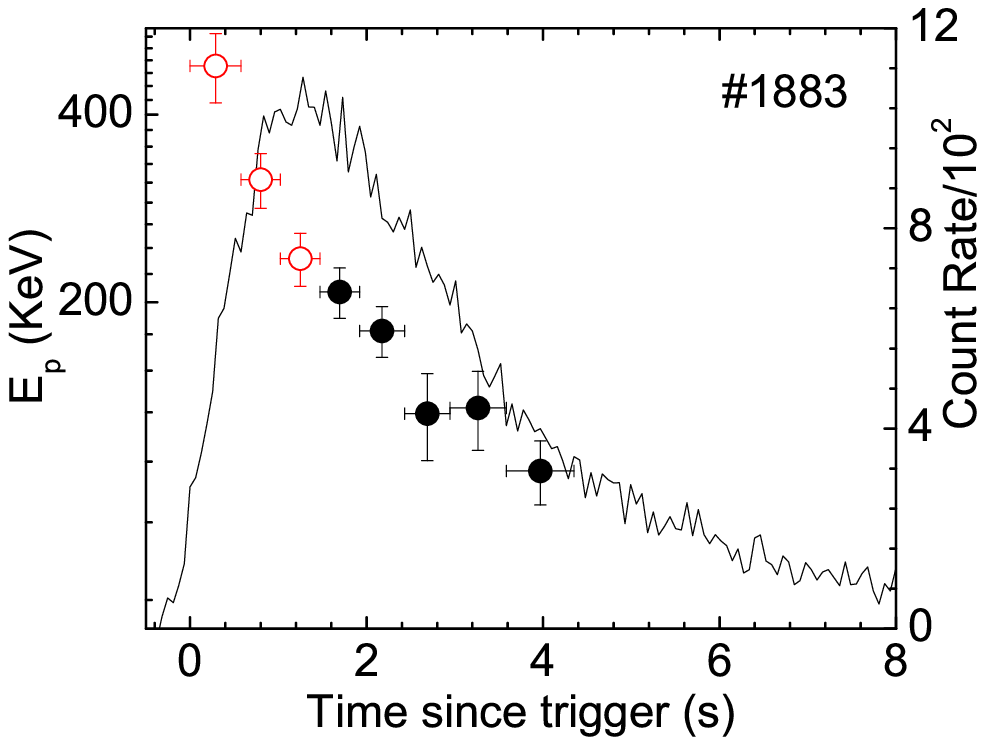}
\includegraphics[scale=0.38]{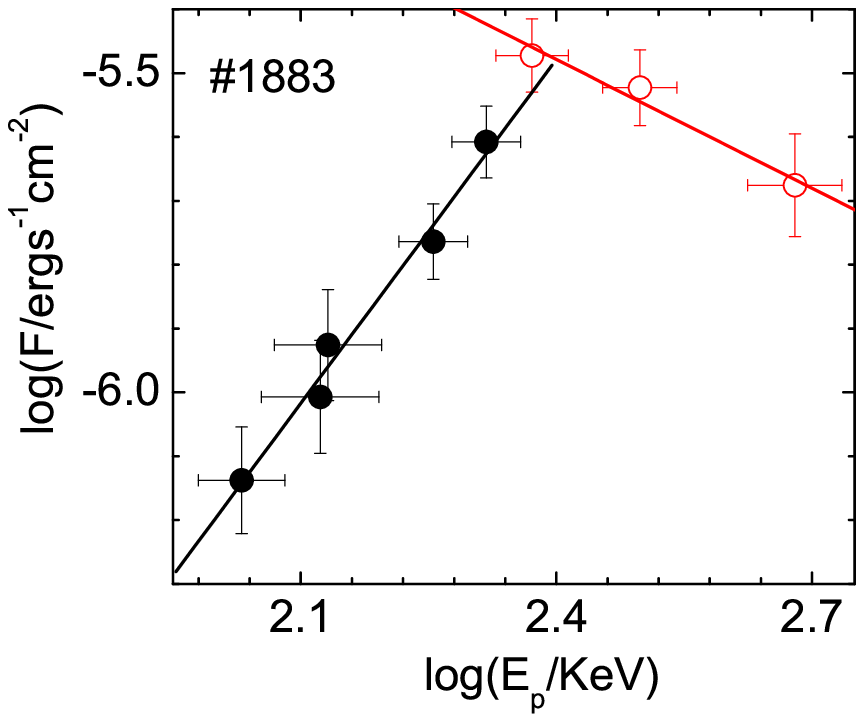}
\includegraphics[scale=0.38]{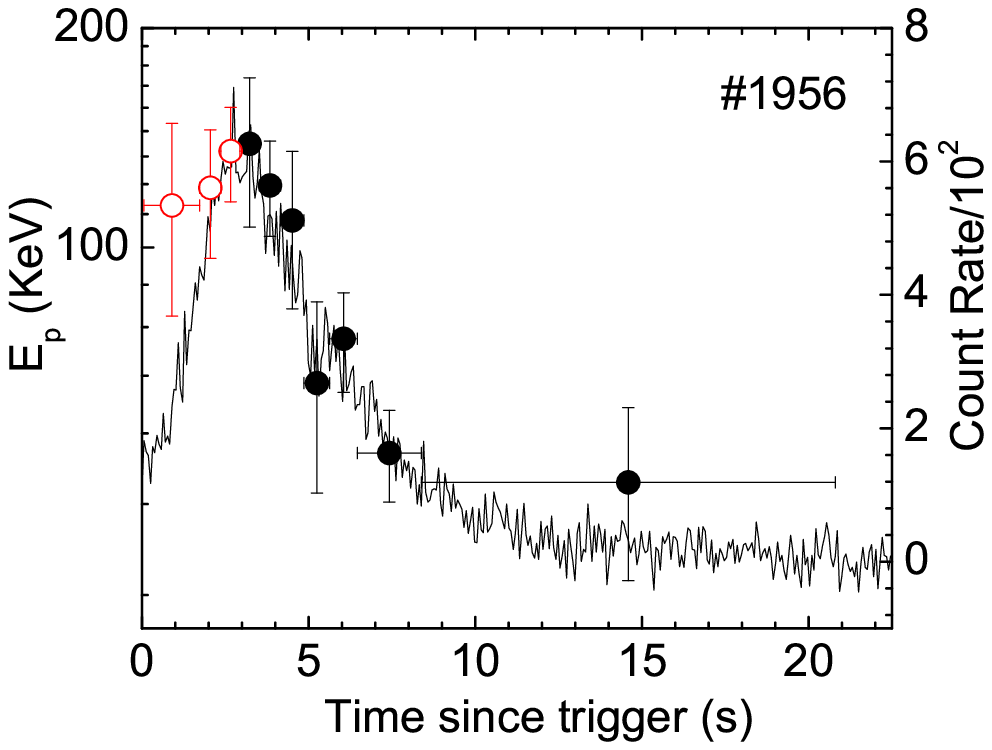}
\includegraphics[scale=0.38]{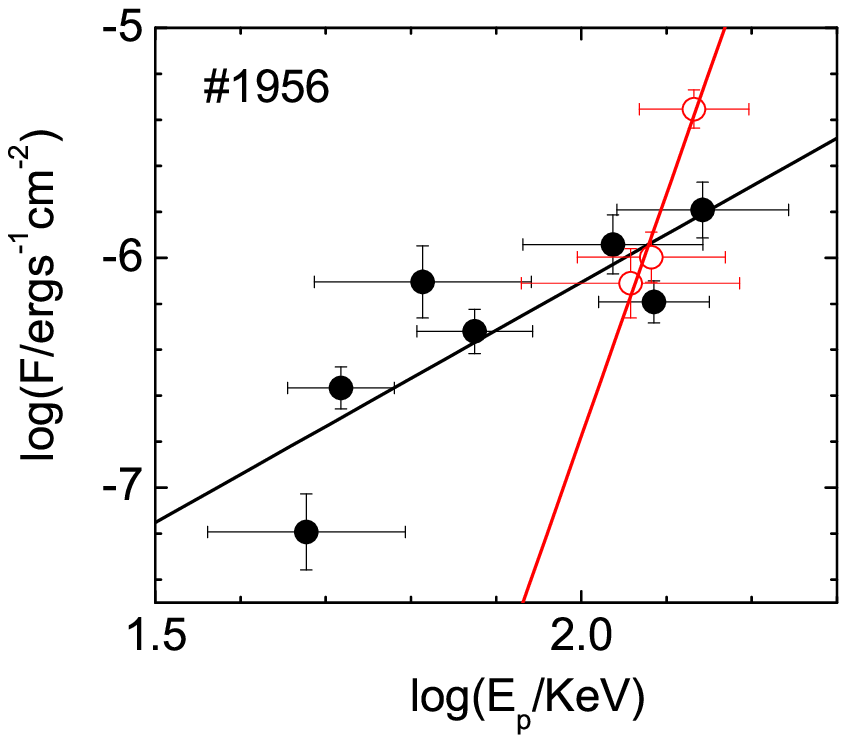}

\includegraphics[scale=0.37]{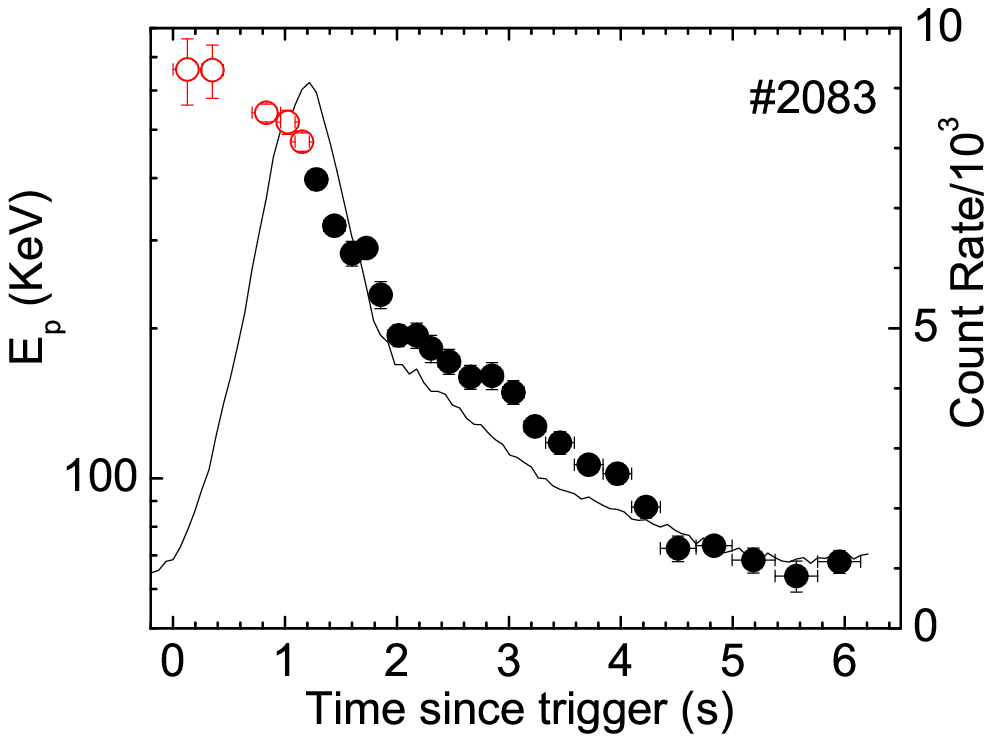}
\includegraphics[scale=0.38]{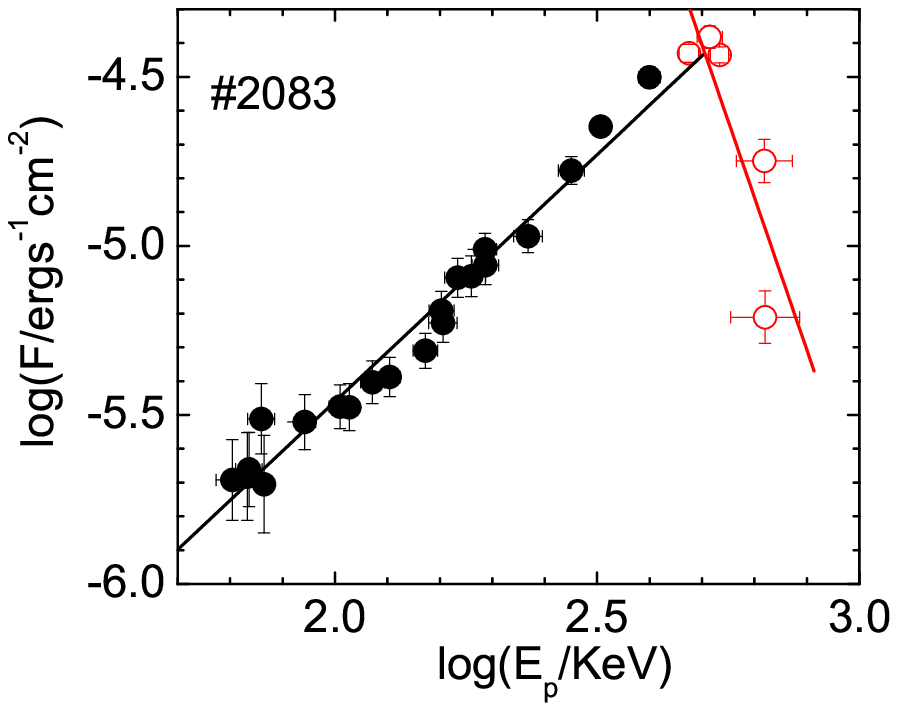}
\includegraphics[scale=0.37]{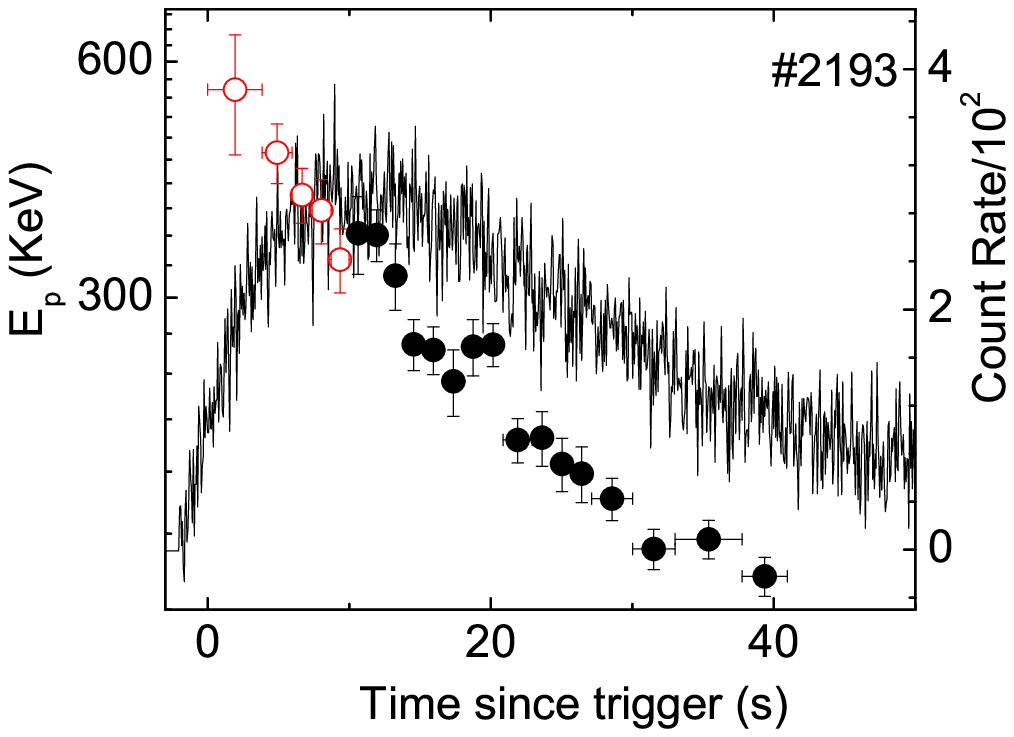}
\includegraphics[scale=0.38]{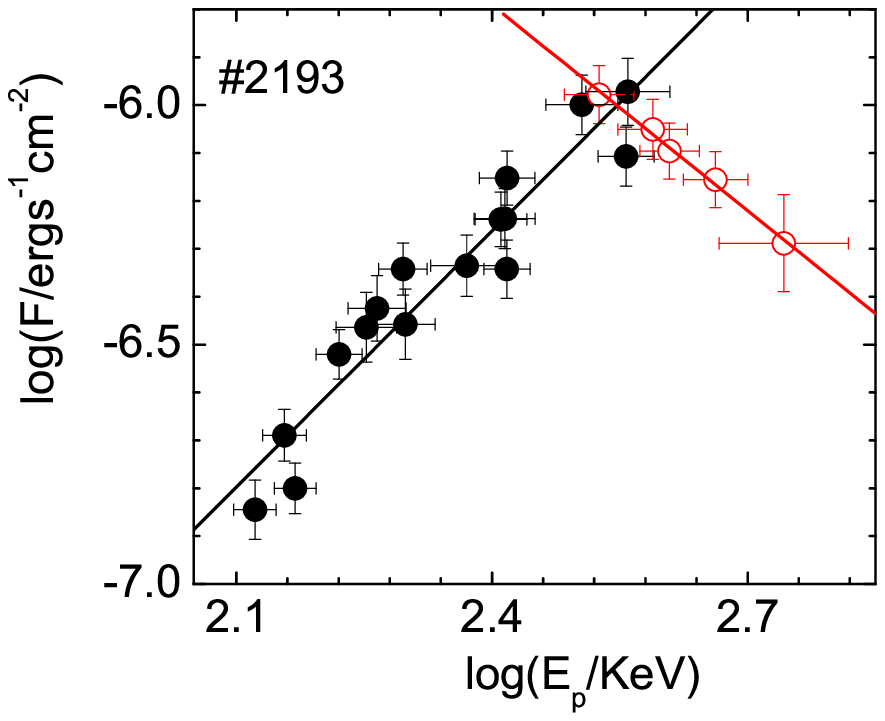}

\includegraphics[scale=0.38]{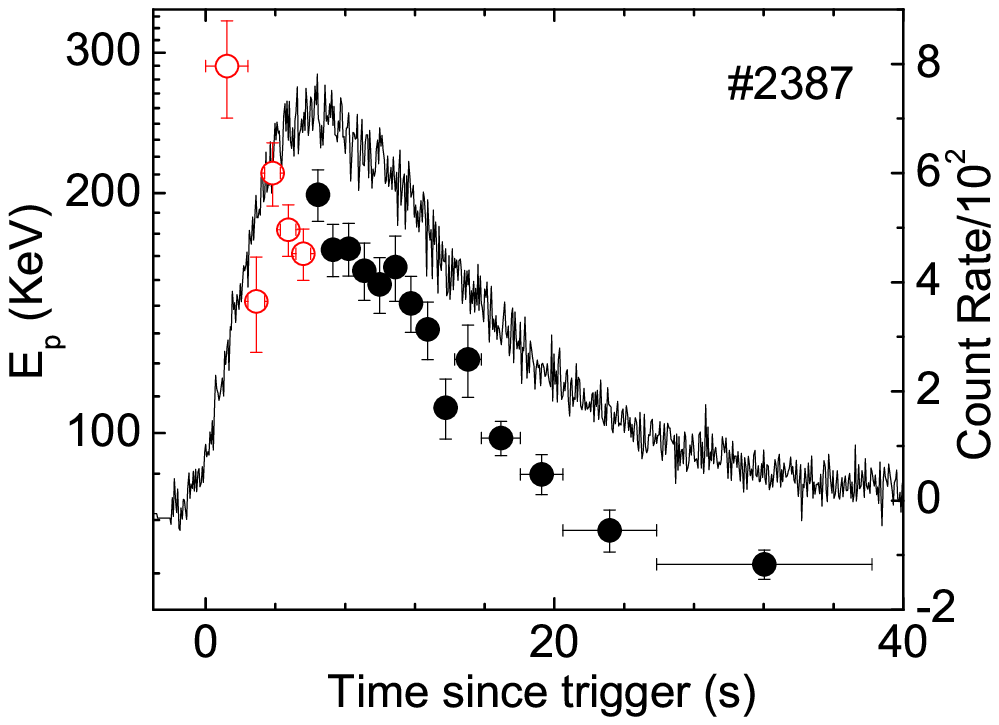}
\includegraphics[scale=0.38]{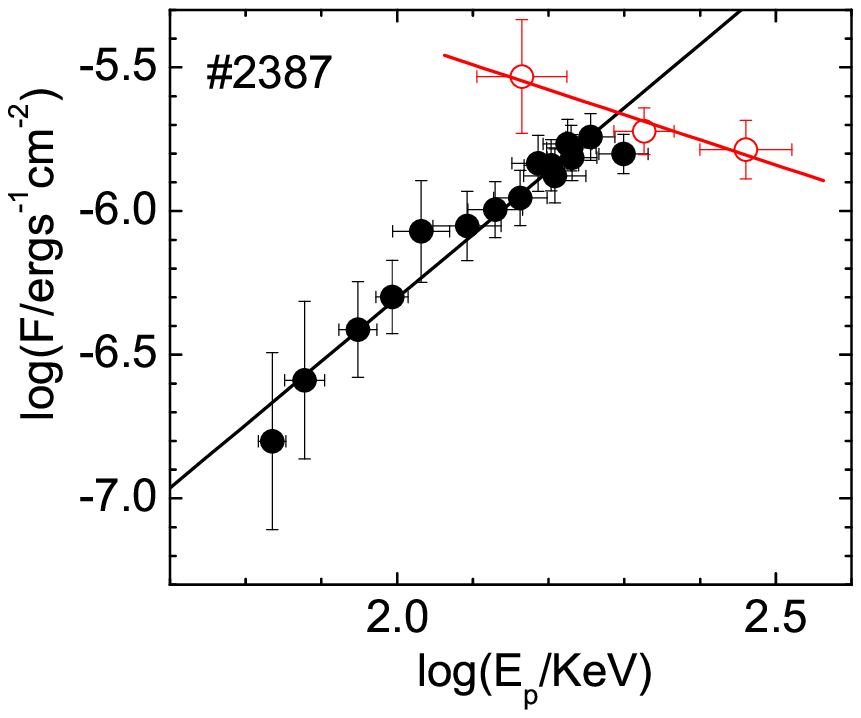}
\includegraphics[scale=0.38]{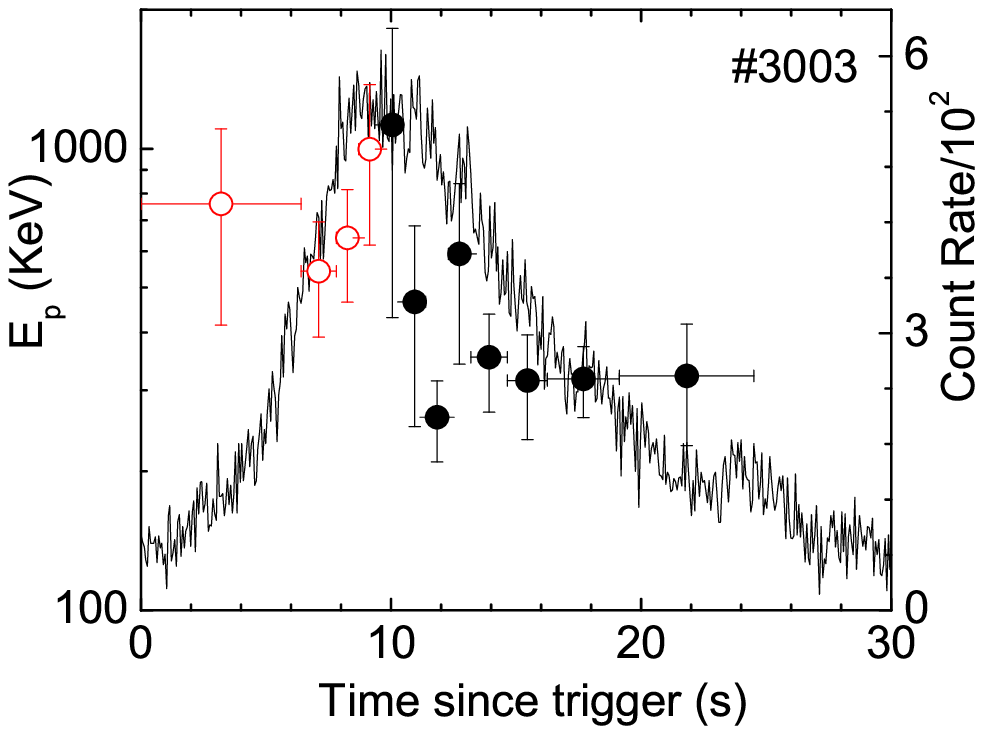}
\includegraphics[scale=0.38]{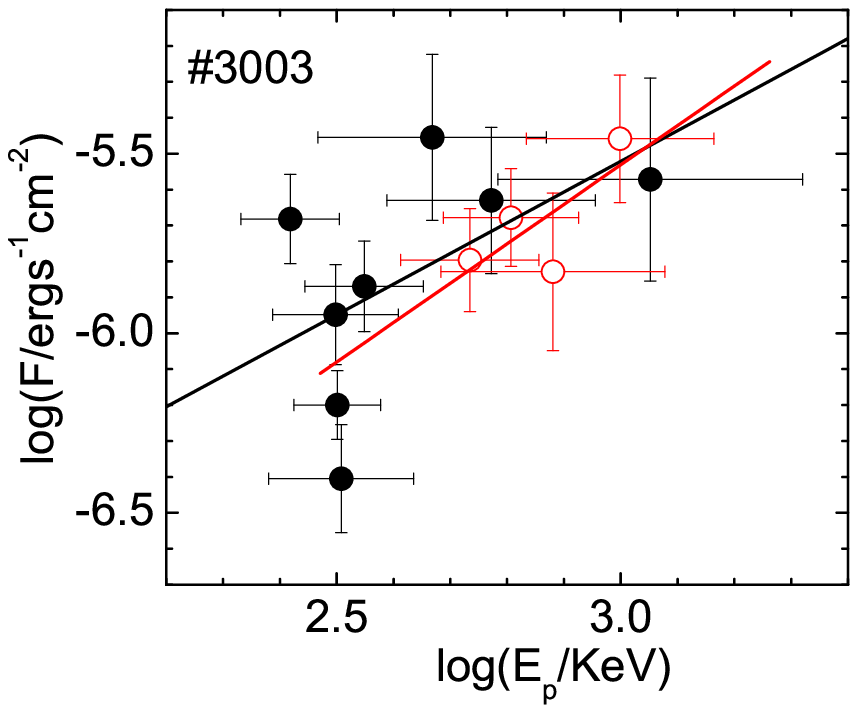}

\includegraphics[scale=0.37]{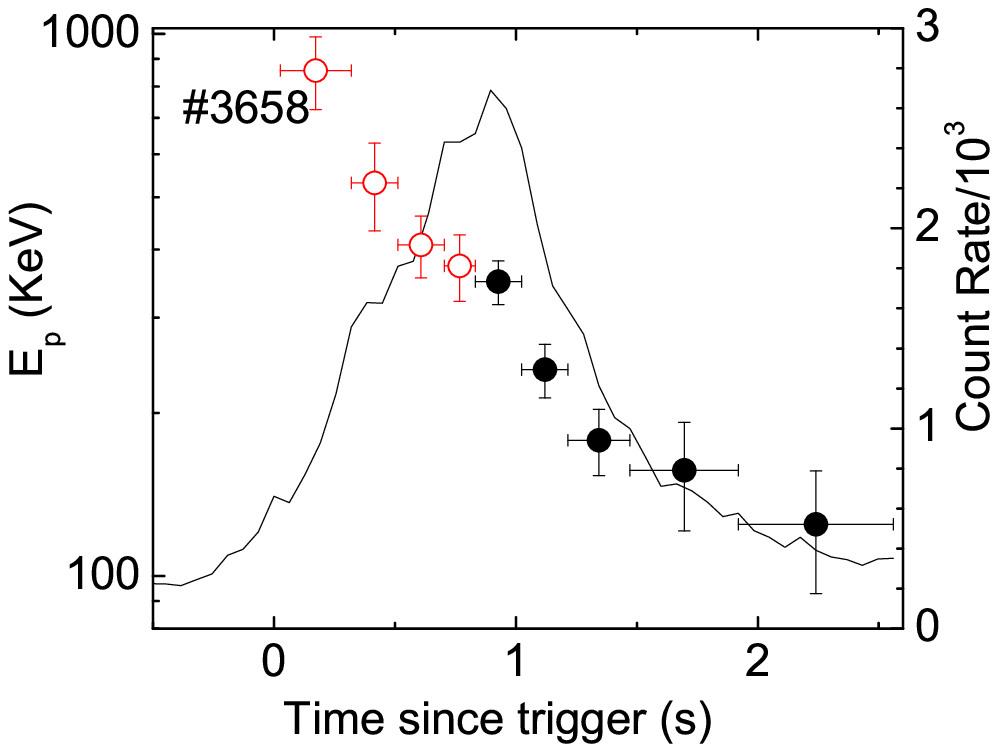}
\includegraphics[scale=0.38]{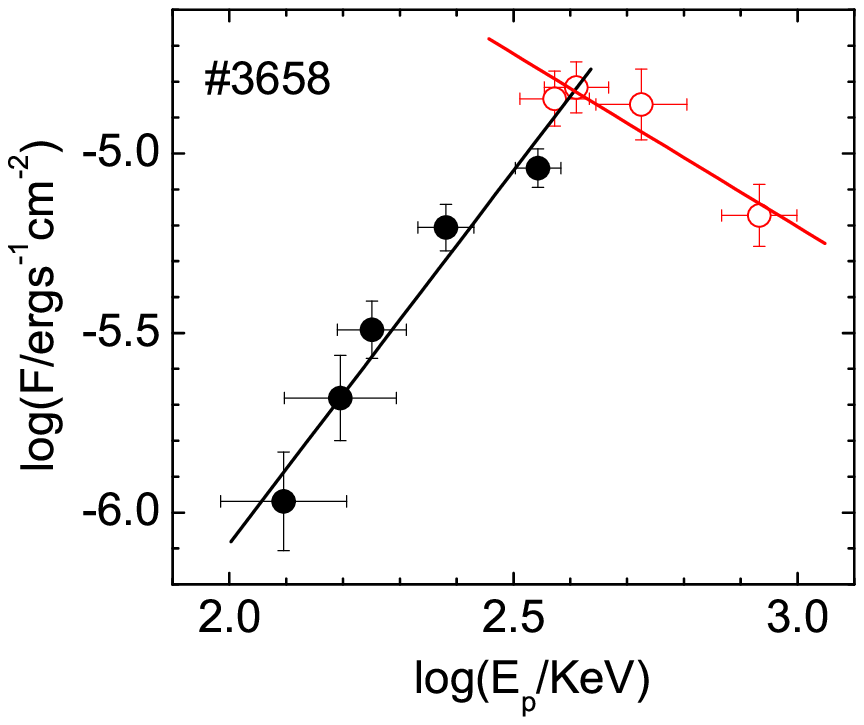}
\includegraphics[scale=0.37]{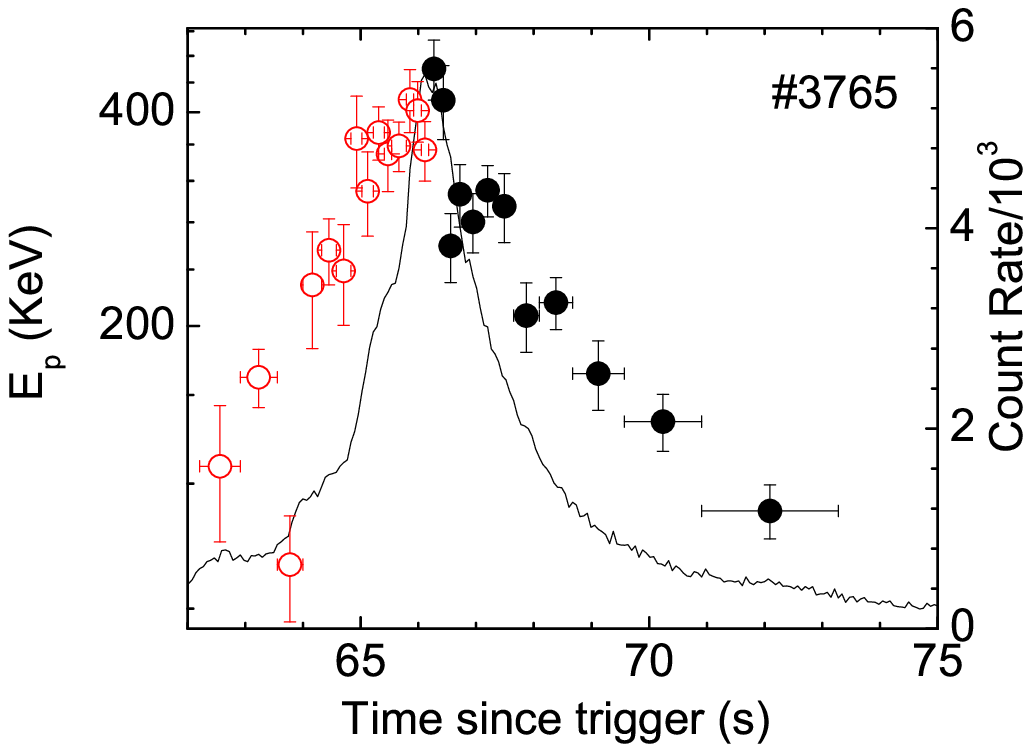}
\includegraphics[scale=0.38]{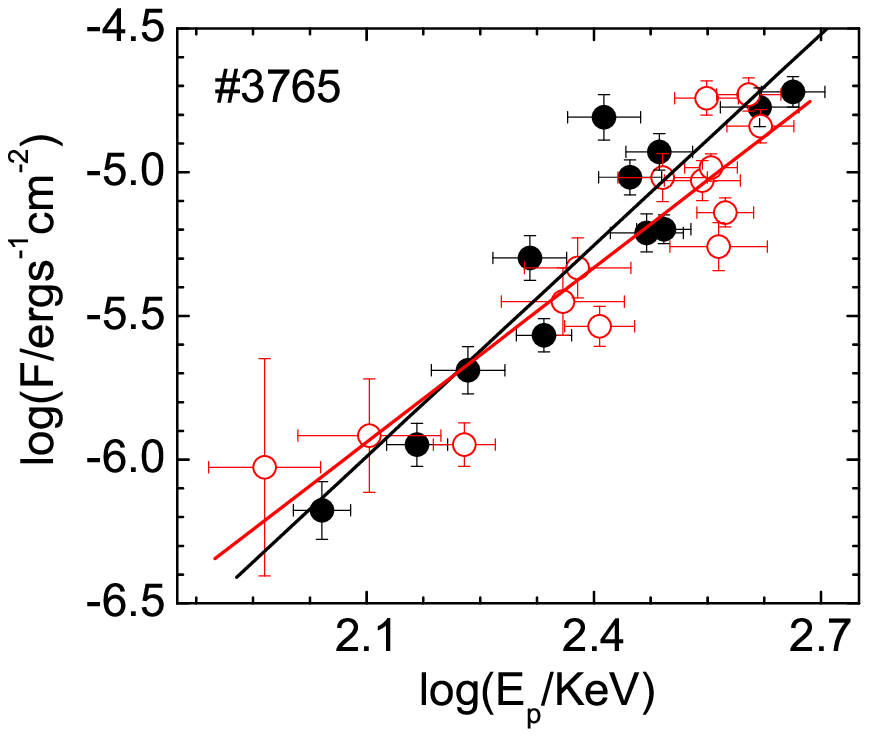}
\caption{Lightcurves with $E_{\rm p}$ evolution (circles) ({\em left
Panels}) and $F-E_{\rm p}$ correlations ({\em right panels}) for the
pulses in our sample. The open circles are for the rising phase and
the solid circles are for decaying phase. Lines are the best fits to
the $F-E_{\rm p}$ correlation. } \label{fig1}
\end{figure*}
 \addtocounter{figure}{-1}

\clearpage

\begin{figure*}
\includegraphics[scale=0.38]{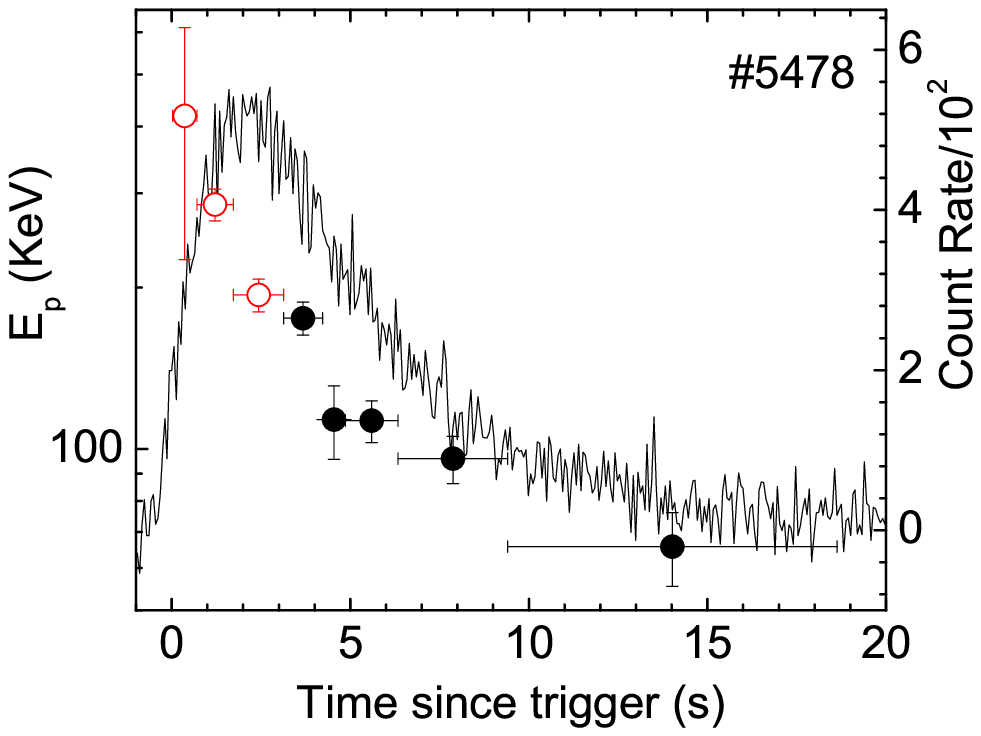}
\includegraphics[scale=0.38]{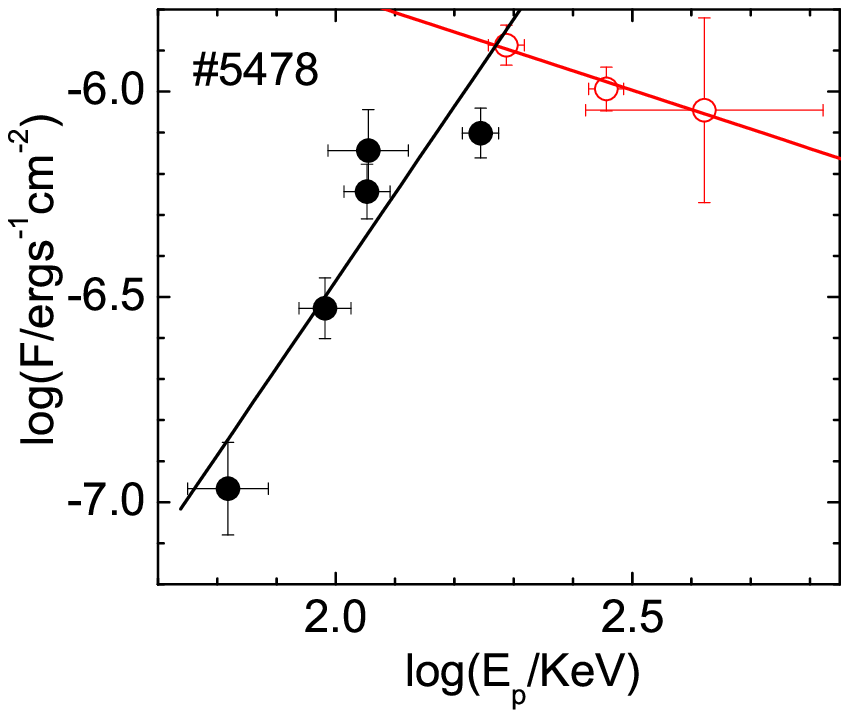}
\includegraphics[scale=0.38]{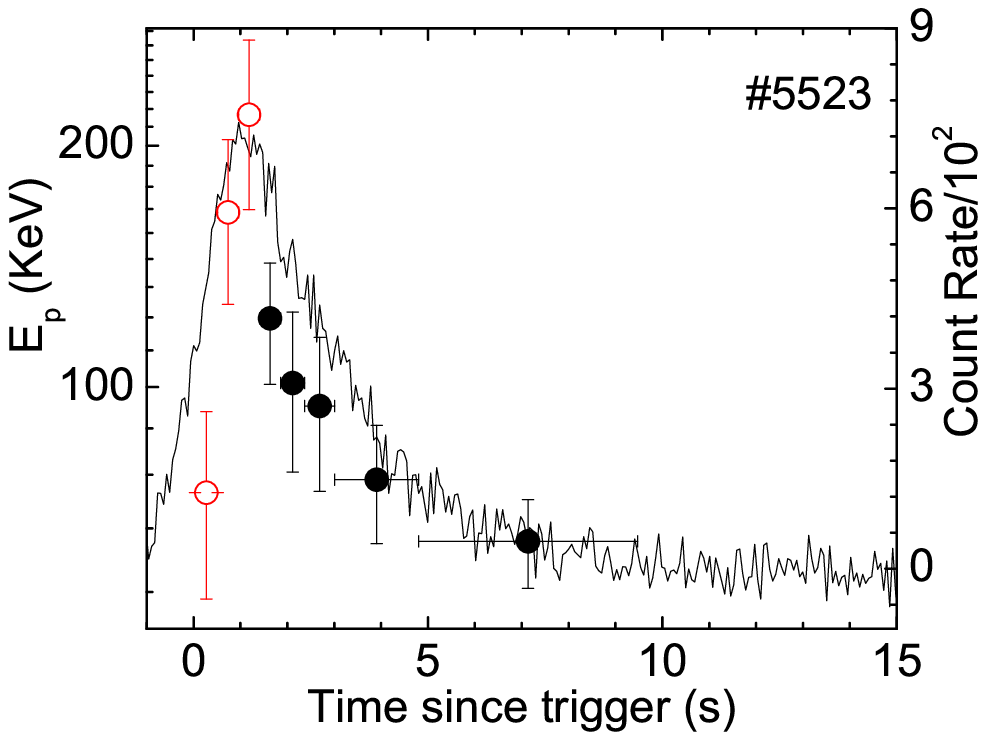}
\includegraphics[scale=0.38]{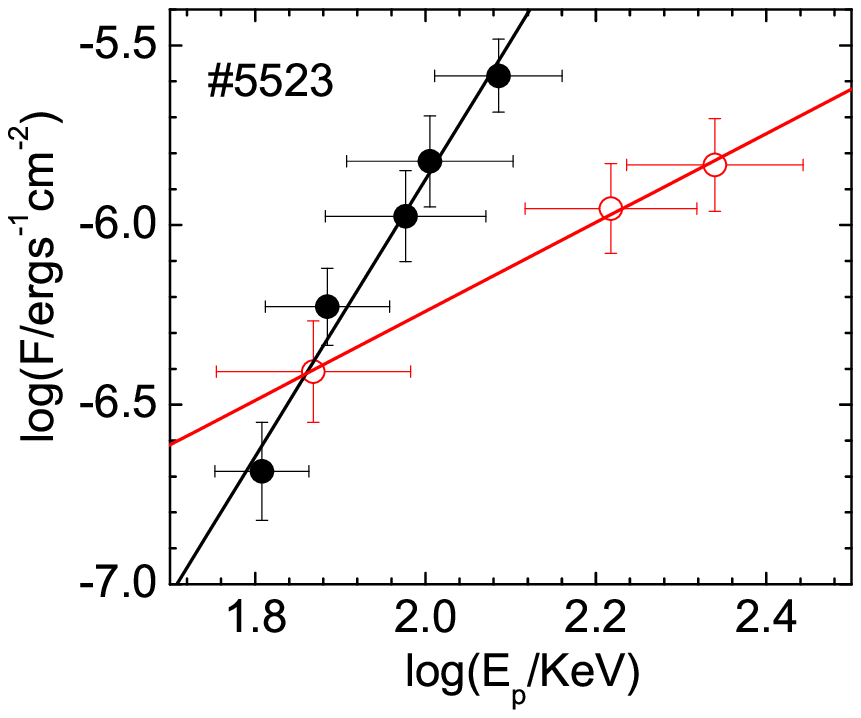}

\includegraphics[scale=0.37]{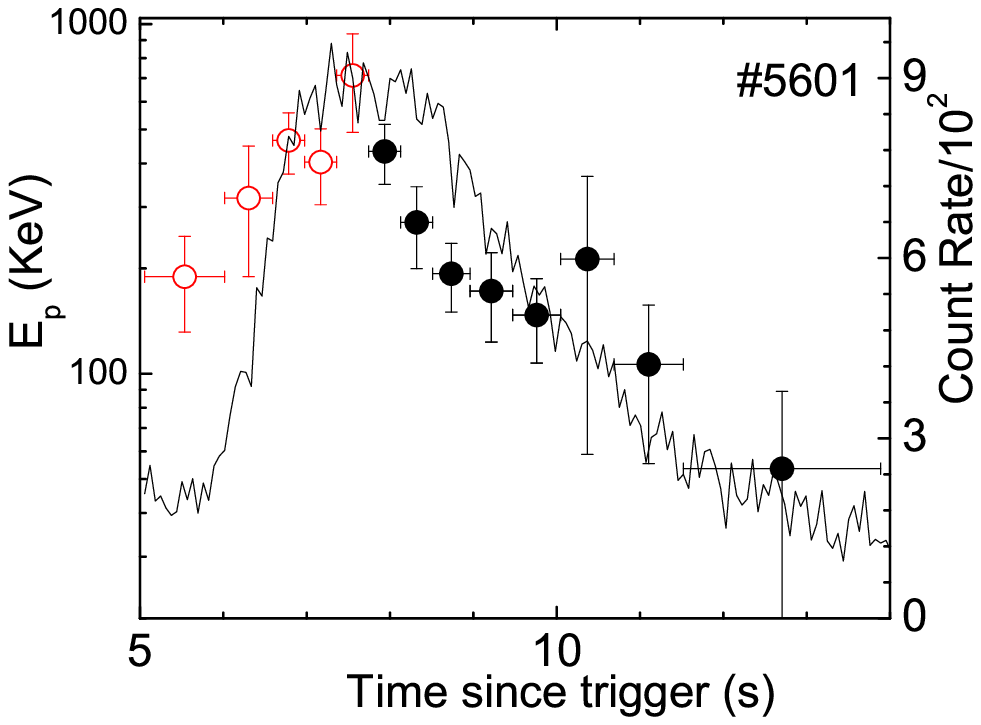}
\includegraphics[scale=0.38]{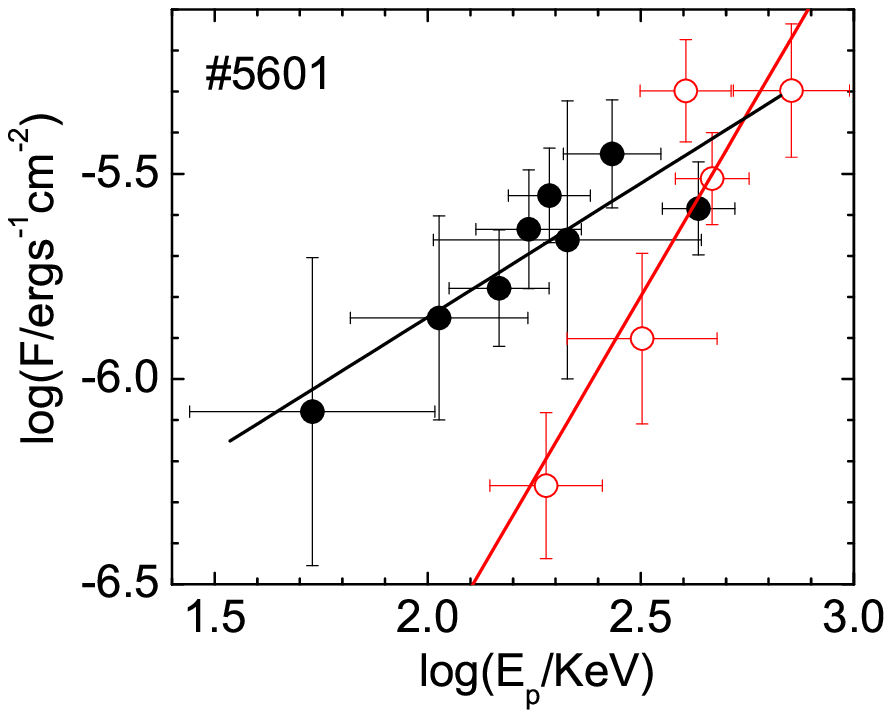}
\includegraphics[scale=0.37]{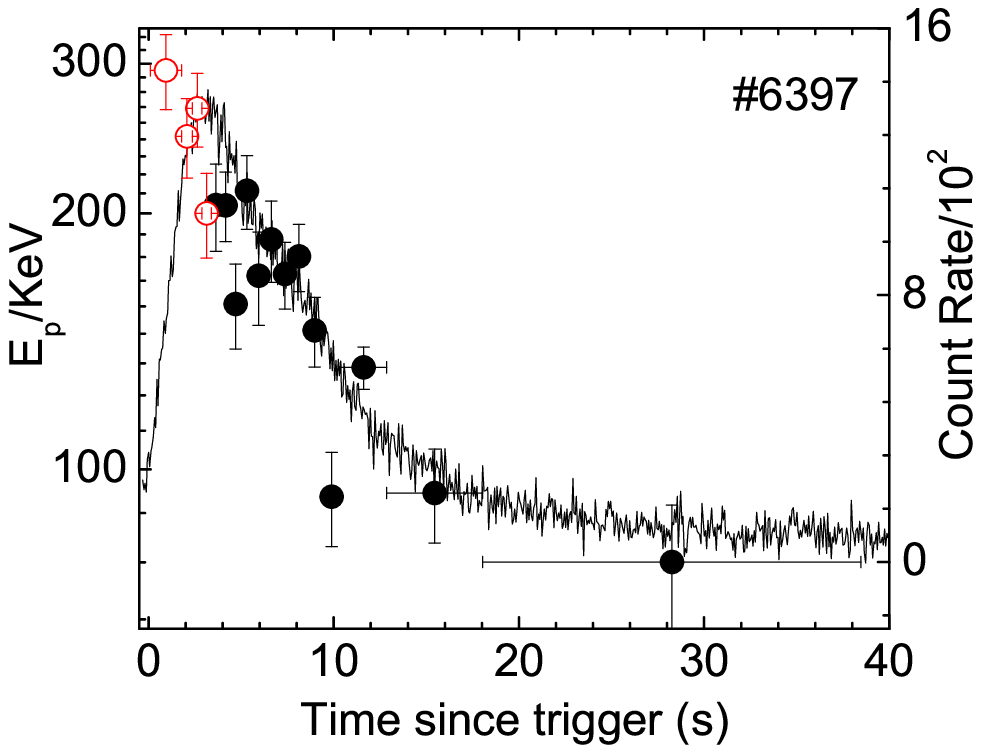}
\includegraphics[scale=0.38]{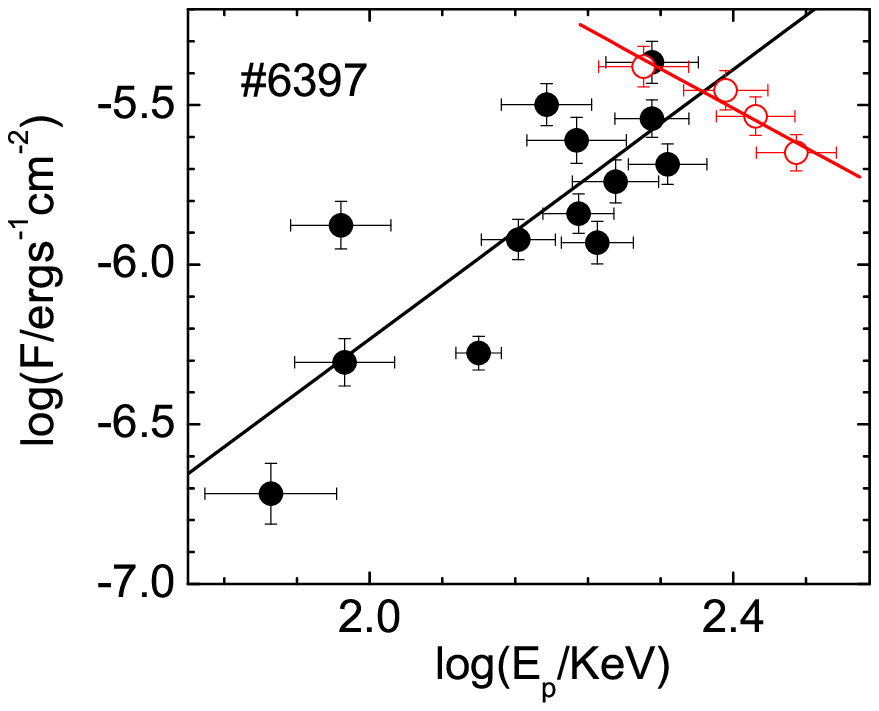}

\includegraphics[scale=0.38]{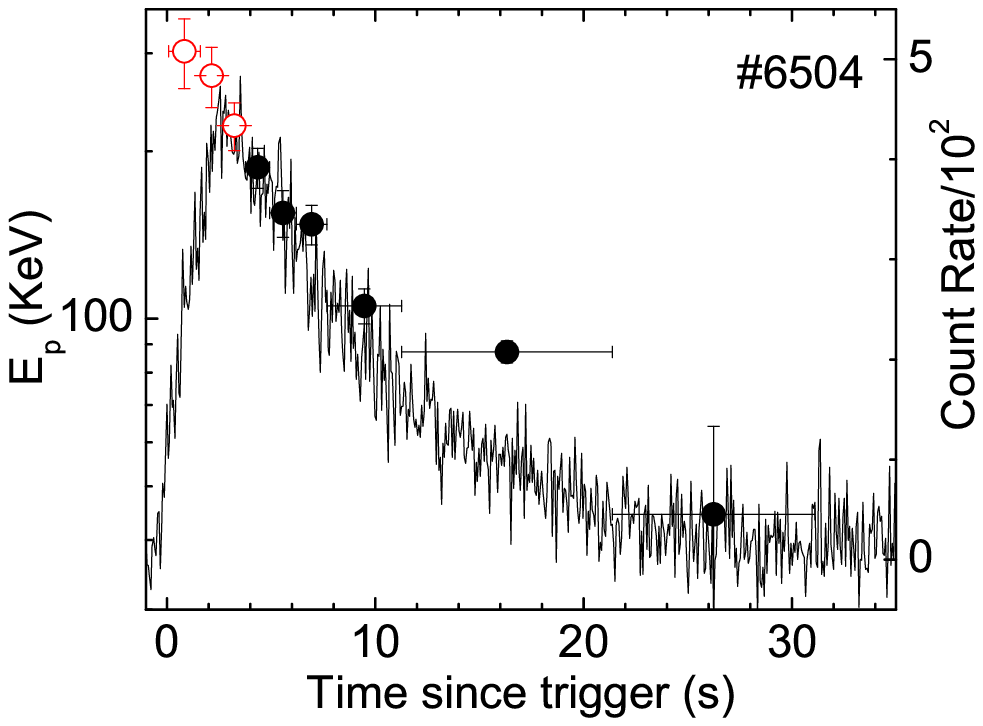}
\includegraphics[scale=0.38]{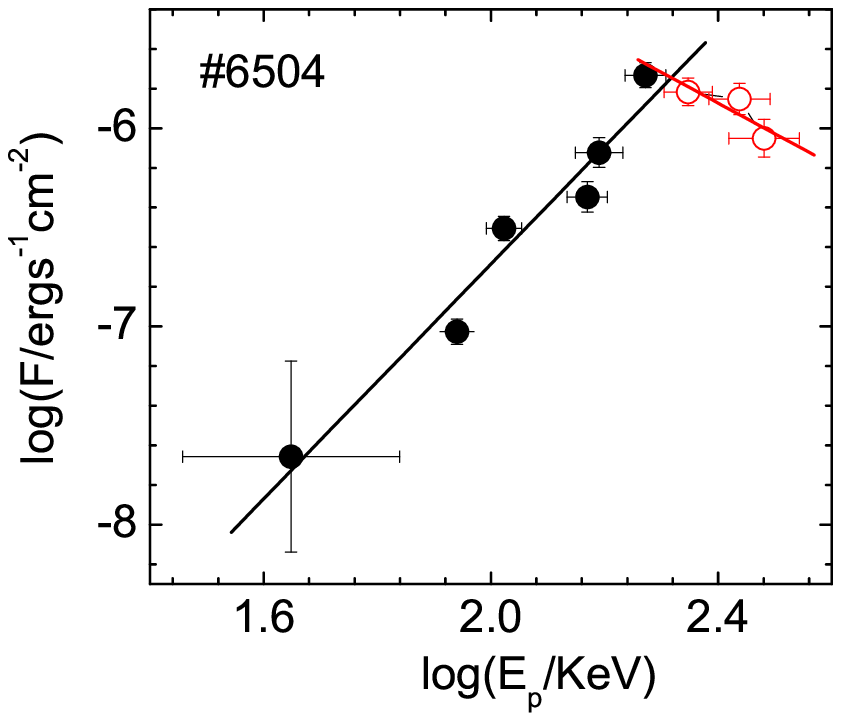}
\includegraphics[scale=0.38]{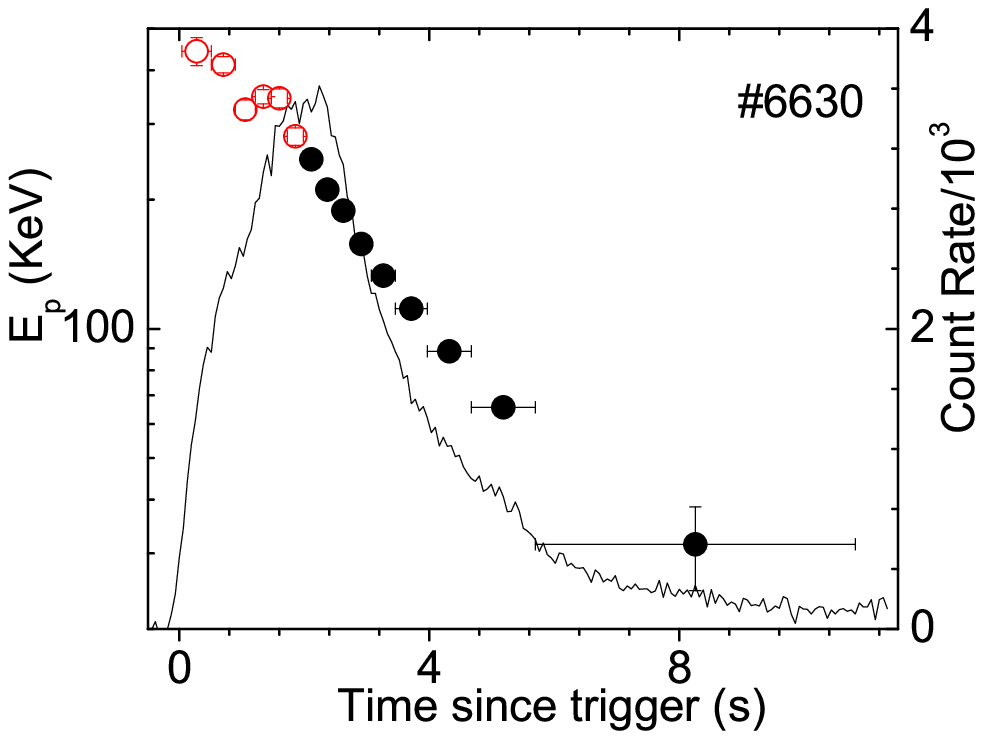}
\includegraphics[scale=0.38]{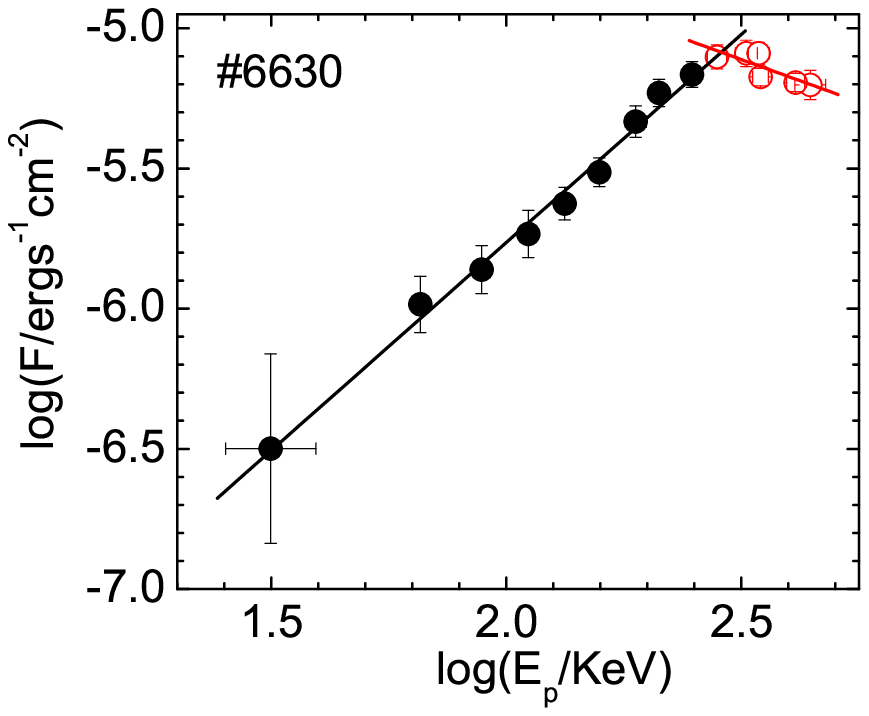}

\includegraphics[scale=0.38]{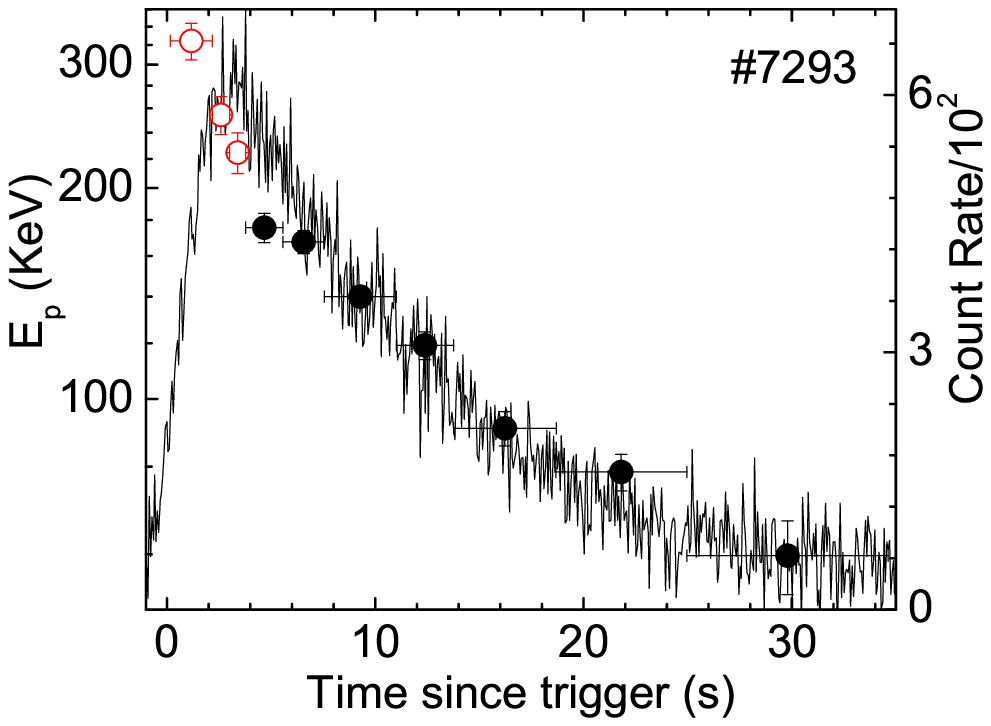}
\includegraphics[scale=0.38]{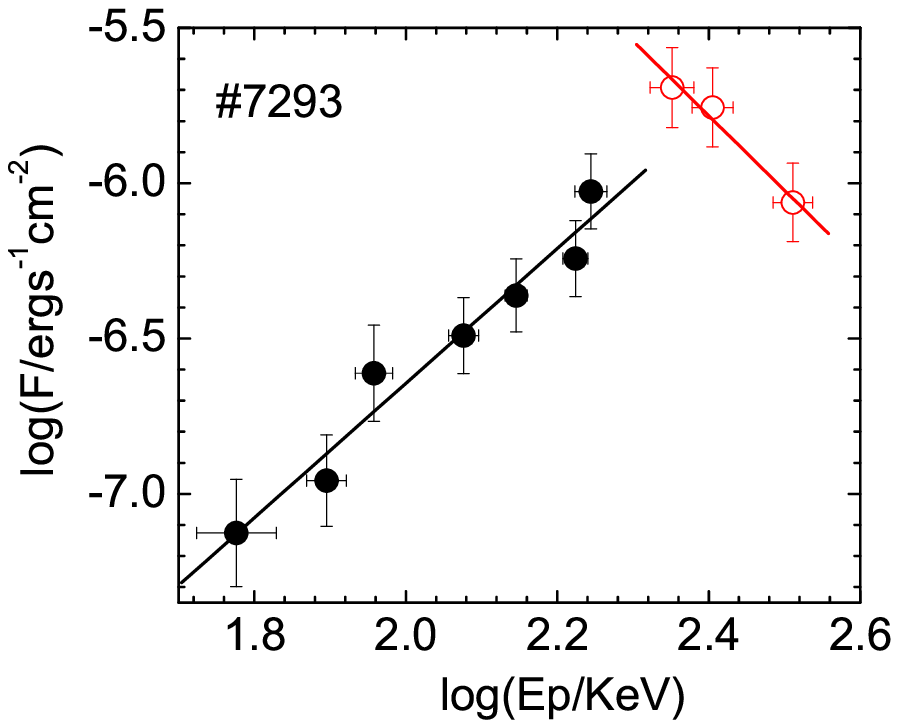}
\includegraphics[scale=0.38]{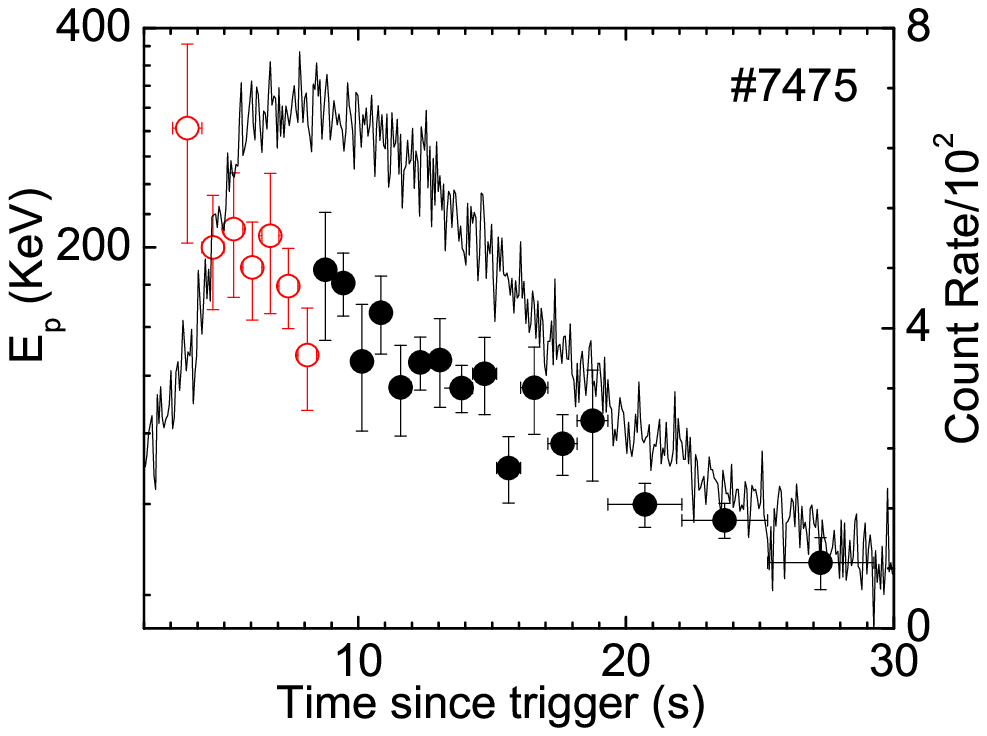}
\includegraphics[scale=0.38]{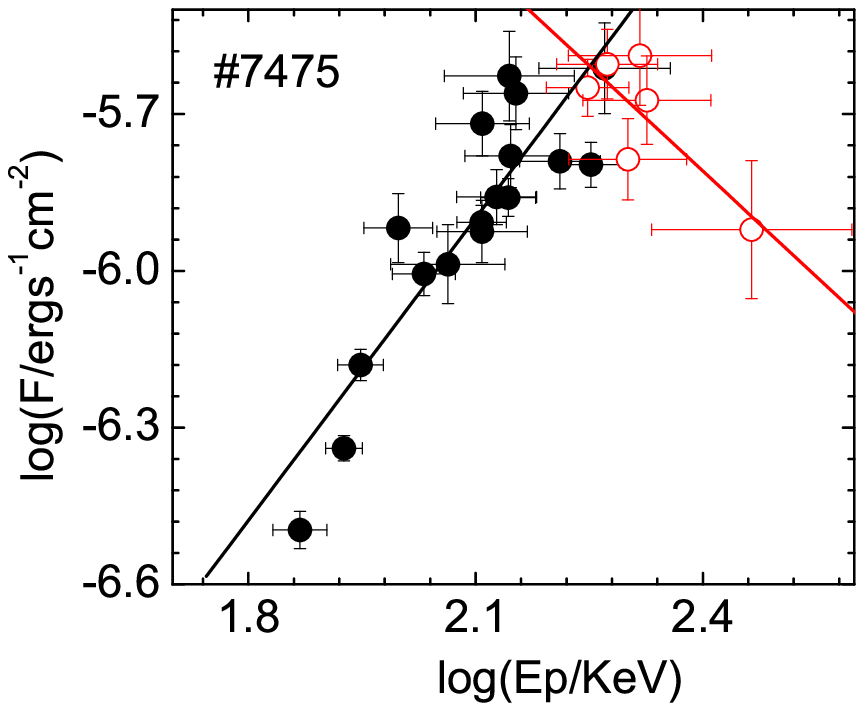}

\includegraphics[scale=0.38]{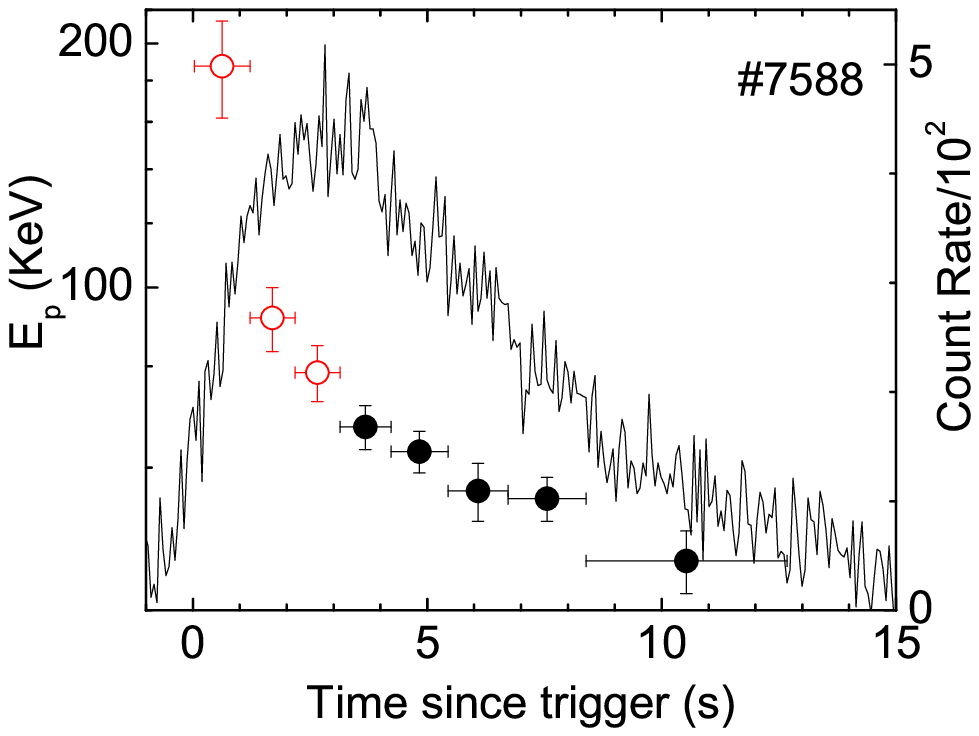}
\includegraphics[scale=0.38]{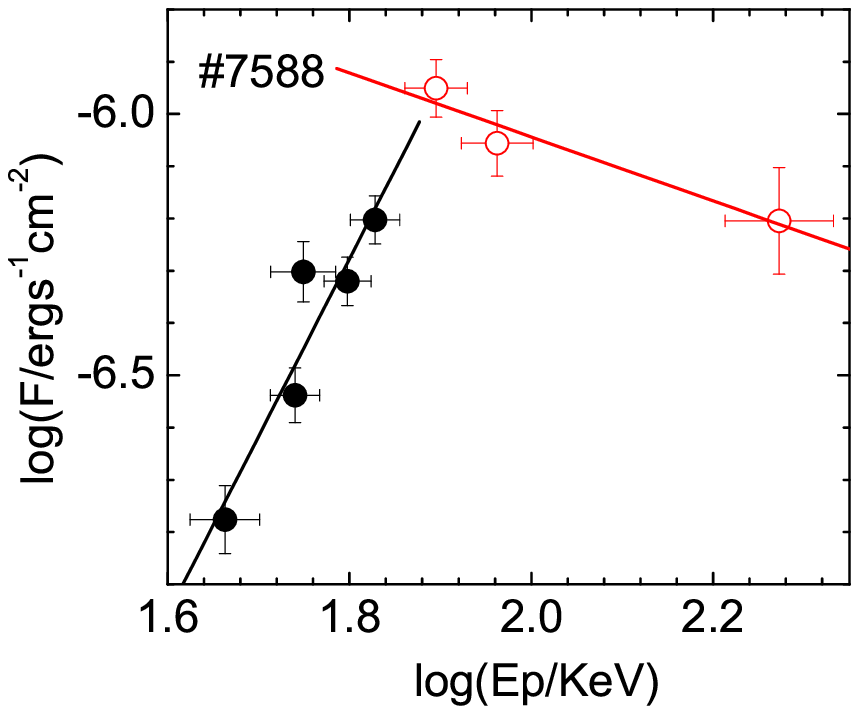}
\includegraphics[scale=0.38]{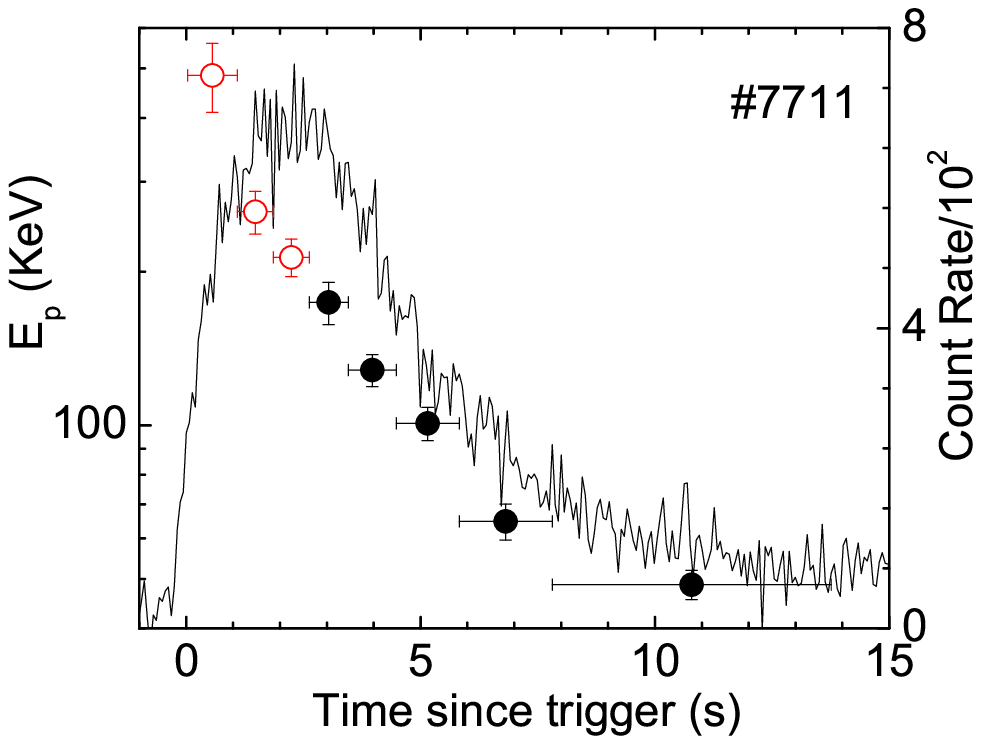}
\includegraphics[scale=0.38]{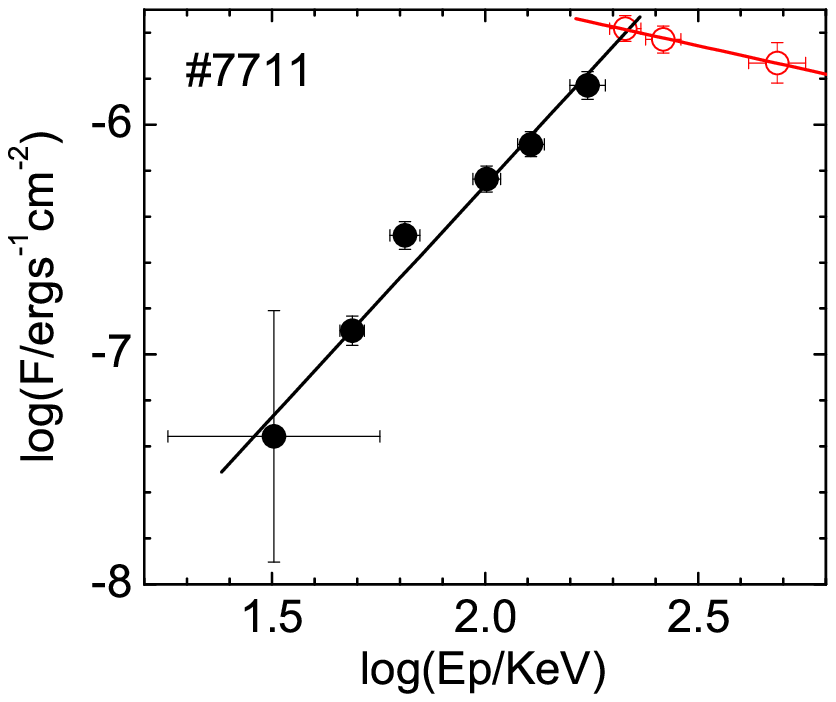}
 \caption{{\it Continued.} }
\end{figure*}
 \addtocounter{figure}{-1}

\begin{figure*}
\includegraphics[scale=0.8]{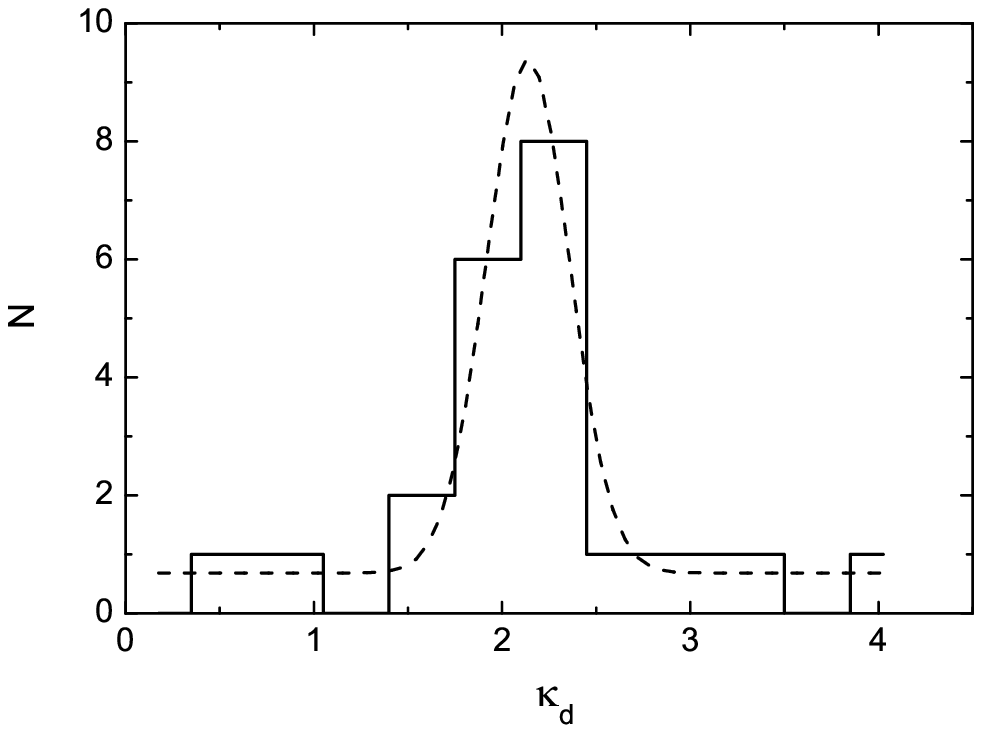}
\caption{Distribution of $\kappa_d$ (the {\em solid} step line) for the pulses
in our sample. The {\em dashed} line is the Gaussian fit, which yields
$\kappa_d=2.14\pm0.44$ ($1\sigma$). }\label{kappad}
\end{figure*}

\begin{figure*}
\includegraphics[scale=0.8]{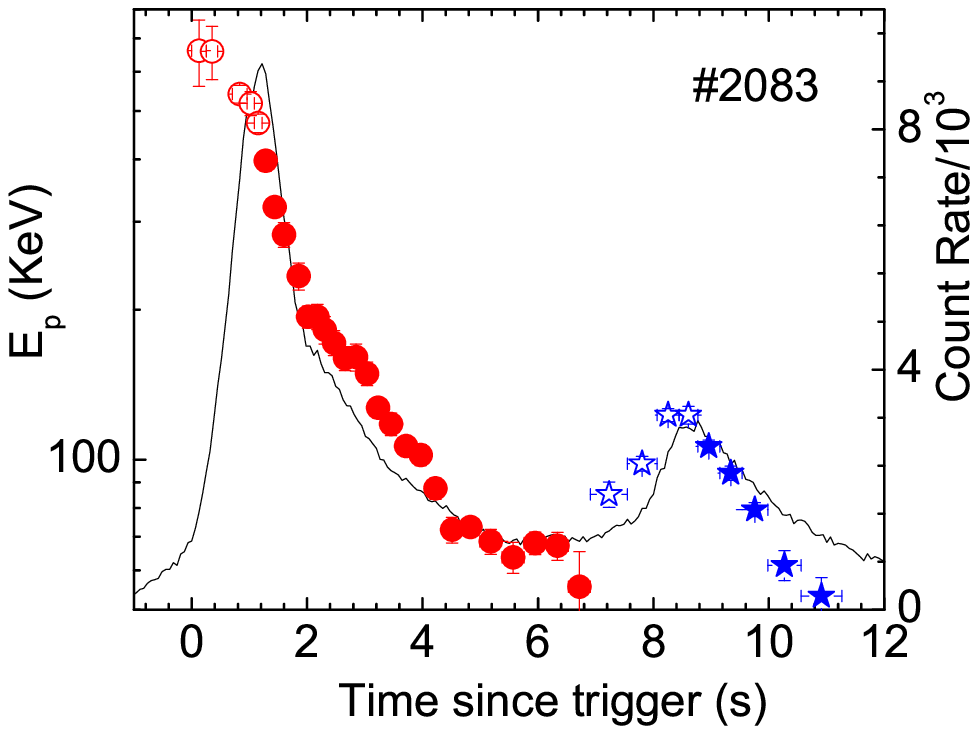}
\caption{The lightcurves and spectral evolution (circles and stars) for two
well-separated pulses in $\#2083$.} \label{two pulses}
\end{figure*}

\begin{figure*}
\includegraphics[scale=0.8]{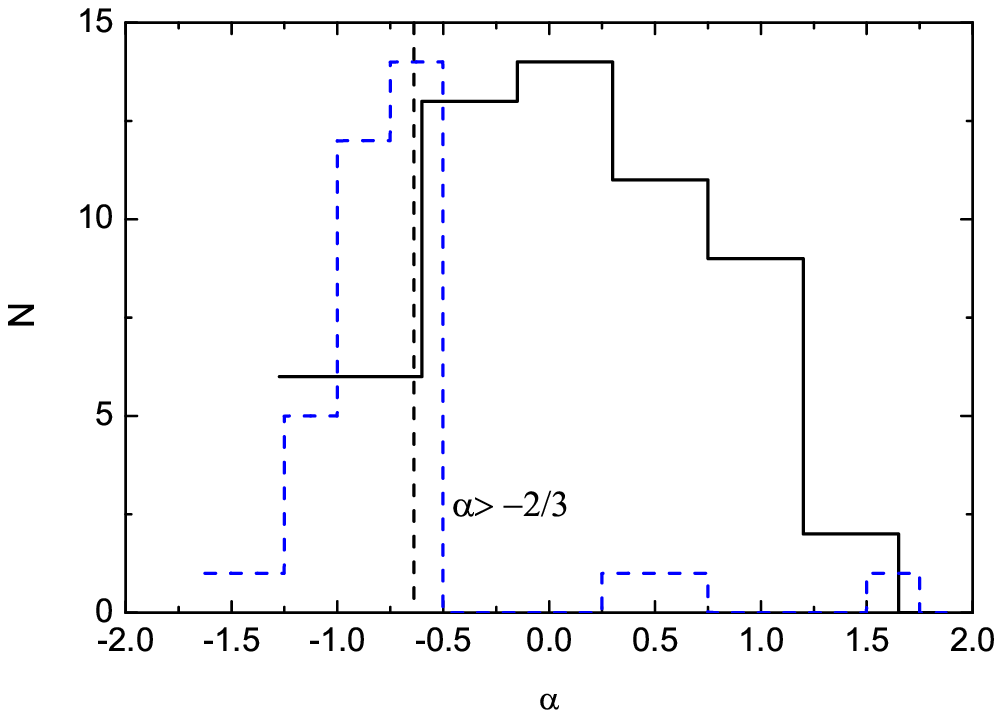}
\caption{Distribution of the low energy photon indices ($\alpha$) in the
spectra whose $E_{\rm p}$ is anti-correlated with the observed fluxes ({\em
solid} step line). The $\alpha$ distribution of the spectra whose $E_{\rm p}$
is correlated with the observed fluxes (dashed line) in the same pulses are
also shown for comparison. The death line ($\alpha=-2/3$) of the synchrotron
radiation is shown with a vertical dashed line.} \label{alpha}
\end{figure*}


\begin{figure*}
\centering
\includegraphics[scale=0.8]{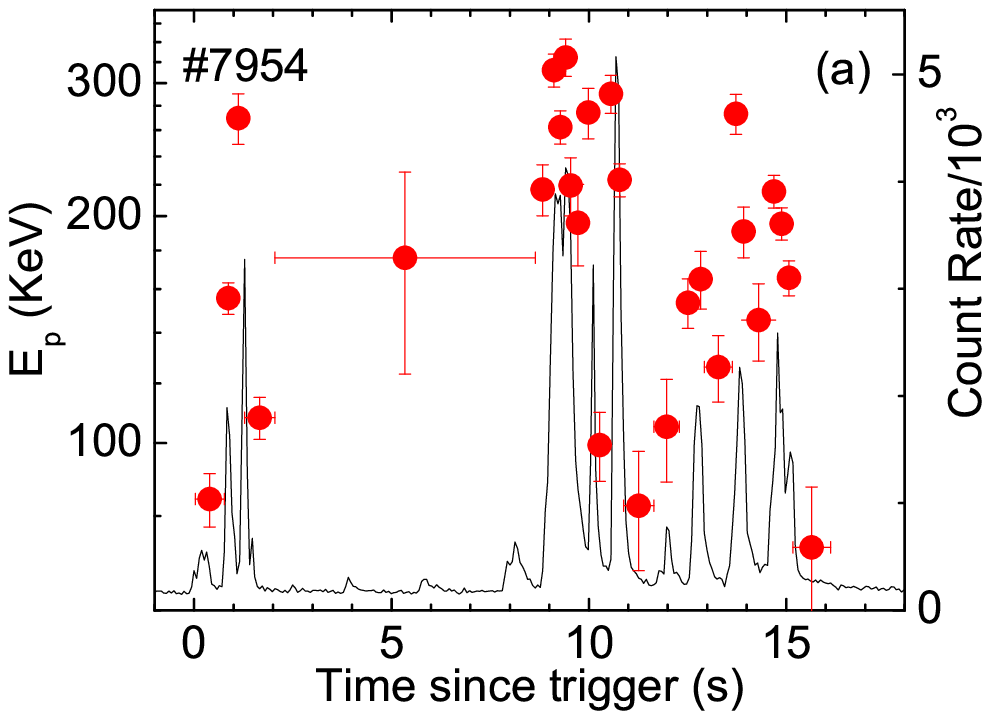}
\includegraphics[scale=0.8]{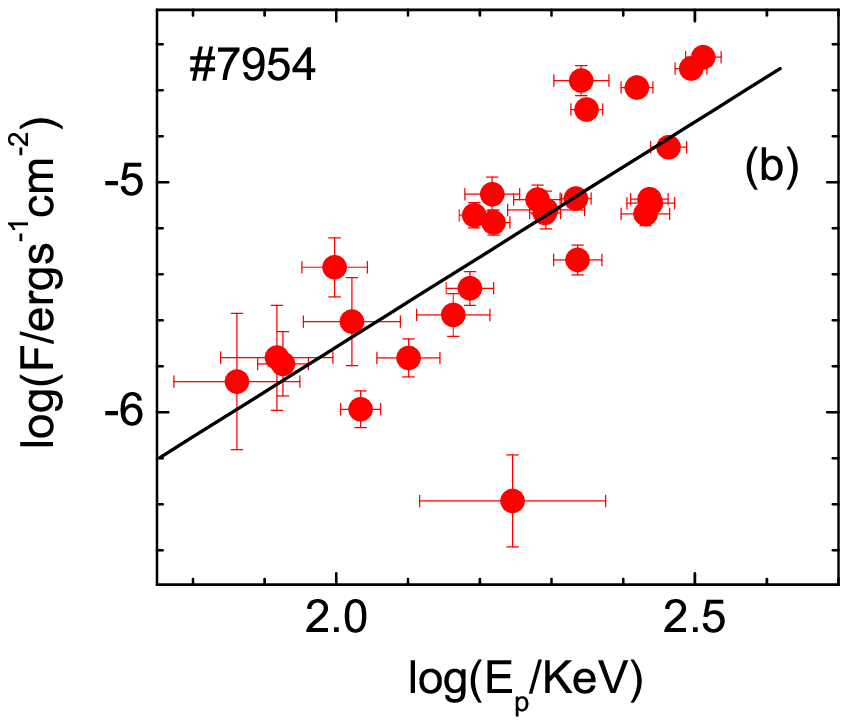}

\caption{Example ($\#7954$) of time-resolved spectra analysis for multi-pulses
GRBs and the $F-E_{\rm p}$ correlation within the GRB. The symbols and line
style are the same as Fig. \ref{fig1}.} \label{Multiple}
\end{figure*}

\clearpage
\begin{center}
\begin{deluxetable}{ccccccc}  
\tablecaption{The power-law indices of the $F\propto E_{\rm
p}^{\kappa_r(\kappa_d)}$ relation for the pulses in our sample. $N$ and $r$ are
the number of the data points and the correlation coefficient,
respectively.\label{tbl}} \tablewidth{0pt} \tablehead{\colhead{Trigger$\#$ } &
\colhead{$\kappa_r$} & \colhead{$r$} & \colhead{N} & \colhead{$\kappa_d$} &
\colhead{$r$} & \colhead{N} } \startdata
       647 &     -2.137$\pm$0.321 &     -0.988 &          3 &      2.775$\pm$0.049 &      0.999 &          3 \\
       973 &     -0.297$\pm$0.150 &     -0.662 &          7 &      2.118$\pm$0.422 &      0.791 &         17 \\
      1733 &      3.877$\pm$1.506 &      0.932 &          3 &      2.176$\pm$0.758 &      0.896 &          4 \\
      1883 &     -0.671$\pm$0.121 &     -0.983 &          3 &      1.796$\pm$0.141 &       0.99 &          5 \\
      1956 &      10.55$\pm$2.011 &      0.982 &          3 &      2.09$\pm$0.643 &      0.823 &          7 \\
      2083 &     -4.515$\pm$1.712 &     -0.835 &          5 &      1.461$\pm$0.060 &      0.983 &         21 \\
      2193 &     -1.432$\pm$0.068 &     -0.996 &          5 &      1.783$\pm$0.148 &      0.954 &         16 \\
      2387 &     -0.871$\pm$0.201 &     -0.974 &          3 &      2.205$\pm$0.137 &      0.973 &         16 \\
      3003 &      1.097$\pm$0.707 &      0.738 &          4 &      0.854$\pm$0.544 &      0.539 &          8 \\
      3658 &     -0.963$\pm$0.253 &     -0.937 &          4 &      2.076$\pm$0.272 &      0.975 &          5 \\
      3765 &     2.0248$\pm$0.249 &      0.919 &         14 &      2.449$\pm$0.299 &      0.932 &         12 \\
      5478 &     -0.473$\pm$0.092 &     -0.981 &          3 &      2.123$\pm$0.569 &      0.906 &          5 \\
      5523 &      1.237$\pm$0.062 &      0.998 &          3 &      3.857$\pm$0.339 &      0.988 &          5 \\
      5601 &      1.779$\pm$0.491 &      0.902 &          5 &      0.651$\pm$0.135 &       0.89 &          8 \\
      6397 &      -1.55$\pm$0.341 &     -0.954 &          4 &      2.110$\pm$0.465 &      0.807 &         13 \\
      6504 &     -1.547$\pm$1.01 &     -0.837 &          3 &       2.961$\pm$ 0.300&      0.979 &          6 \\
      6630 &     -0.606$\pm$0.217 &     -0.812 &          6 &      1.483$\pm$0.051 &      0.995 &          9 \\
      7293 &     -2.401$\pm$0.422 &     -0.984 &          3 &      2.166$\pm$0.204 &      0.978 &          7 \\
      7475 &     -1.345$\pm$0.506 &     -0.799 &          6 &      1.927$\pm$0.26 &      0.885 &         17 \\
      7588 &     -0.612$\pm$0.161 &     -0.966 &          3 &      3.399$\pm$0.765 &      0.931 &          5 \\
      7711 &     -0.410$\pm$0.032 &     -0.997 &          3 &      2.015$\pm$0.16 &      0.987 &          6 \\
\enddata

\end{deluxetable}
\end{center}

\begin{center}
\begin{deluxetable}{ccccccc}
\tablecaption{Online Material: Spectral fits with the Band function.
The flux is in the $30-10^4$ keV band. \label{tb2}}
\tablehead{\colhead{Trigger$\#$ } & \colhead{$t_{\rm start}$(s)} &
\colhead{$t_{\rm end}$(s)} & \colhead{$E_{p}$(keV)} &
\colhead{$\alpha$} & \colhead{$\beta$} & \colhead{flux ($10^{-6}{\rm
erg/cm}^{2} {\rm s}$)} } \startdata
       647 &          0 &      1.472 &     276.06 $\pm$  21.61 &       0.53 $\pm$   0.24 &      -2.65 $\pm$   0.28 &       1.47 $\pm$   0.24 \\
       647 &      1.472 &       2.56 &     228.64 $\pm$  11.35 &       0.74 $\pm$    0.2 &      -2.86 $\pm$   0.23 &       2.09 $\pm$   0.26 \\
       647 &       2.56 &      3.456 &     219.13 $\pm$   7.88 &       0.75 $\pm$   0.16 &       -3.4 $\pm$   0.34 &       2.48 $\pm$   0.24 \\
       647 &      3.456 &      4.416 &      176.7 $\pm$  7.02  &       0.63 $\pm$   0.19 &      -3.16 $\pm$   0.26 &        1.9 $\pm$   0.24 \\
       647 &      4.416 &      5.568 &     150.05 $\pm$   6.32 &       0.44 $\pm$   0.22 &      -3.33 $\pm$   0.36 &       1.19 $\pm$   0.19 \\
       647 &      5.568 &      6.848 &     138.34 $\pm$   7.03 &       0.26 $\pm$   0.27 &      -3.14 $\pm$   0.32 &       0.96 $\pm$   0.19 \\
       973 &          0 &      0.576 &     797.06 $\pm$ 435.63 &      -0.99 $\pm$   0.15 &      -1.76 $\pm$   0.29 &       3.93 $\pm$   2.37 \\
       973 &      0.576 &      1.024 &     891.13 $\pm$ 410.27 &      -1.02 $\pm$   0.11 &      -1.82 $\pm$   0.31 &          6 $\pm$   3.01 \\
       973 &      1.024 &      1.408 &     231.82 $\pm$  78.66 &      -0.49 $\pm$   0.35 &      -1.57 $\pm$   0.06 &       8.39 $\pm$   4.54 \\
       973 &      1.408 &      1.792 &     343.63 $\pm$  68.95 &      -0.73 $\pm$   0.15 &      -1.93 $\pm$   0.13 &       5.75 $\pm$   1.55 \\
       973 &      1.792 &      2.112 &     530.62 $\pm$ 172.22 &      -1.03 $\pm$   0.12 &      -1.89 $\pm$   0.19 &       7.21 $\pm$   2.71 \\
       973 &      2.112 &      2.432 &     271.79 $\pm$  45.74 &       -0.5 $\pm$   0.19 &       -1.9 $\pm$    0.1 &       6.92 $\pm$    1.8 \\
       973 &      2.432 &      2.752 &     315.61 $\pm$  57.94 &      -0.65 $\pm$   0.16 &      -1.95 $\pm$   0.13 &       6.42 $\pm$   1.66 \\
       973 &      2.752 &      3.072 &     346.97 $\pm$  69.75 &       -0.9 $\pm$   0.13 &      -2.05 $\pm$   0.17 &       5.93 $\pm$   1.55 \\
       973 &      3.072 &      3.392 &     366.84 $\pm$ 116.22 &      -1.01 $\pm$   0.15 &      -1.79 $\pm$    0.1 &       8.25 $\pm$   3.23 \\
       973 &      3.392 &      3.712 &     250.95 $\pm$  60.93 &      -0.84 $\pm$    0.2 &      -1.87 $\pm$    0.1 &       6.22 $\pm$   2.19 \\
       973 &      3.712 &      4.032 &     362.95 $\pm$   89.4 &      -1.03 $\pm$   0.13 &      -2.08 $\pm$   0.22 &       4.89 $\pm$   1.51 \\
       973 &      4.032 &      4.416 &     248.49 $\pm$  63.74 &      -0.91 $\pm$    0.2 &      -1.86 $\pm$   0.09 &       5.56 $\pm$   2.03 \\
       973 &      4.416 &        4.8 &     212.21 $\pm$  41.97 &      -0.73 $\pm$   0.22 &      -1.95 $\pm$    0.1 &       4.67 $\pm$    1.5 \\
       973 &        4.8 &      5.184 &     173.57 $\pm$  37.11 &      -0.71 $\pm$    0.3 &      -1.95 $\pm$   0.09 &       4.05 $\pm$   1.58 \\
       973 &      5.184 &      5.568 &     194.21 $\pm$  39.52 &      -0.77 $\pm$   0.25 &      -2.01 $\pm$   0.11 &        3.6 $\pm$   1.24 \\
       973 &      5.568 &      6.016 &     195.96 $\pm$  51.96 &      -1.04 $\pm$   0.24 &      -1.99 $\pm$   0.11 &        3.3 $\pm$   1.33 \\
       973 &      6.016 &      6.464 &     184.43 $\pm$  54.69 &      -0.99 $\pm$   0.29 &      -1.95 $\pm$   0.11 &       3.14 $\pm$   1.47 \\
       973 &      6.464 &      6.912 &     174.99 $\pm$     48 &      -0.87 $\pm$   0.33 &      -1.96 $\pm$   0.11 &       2.86 $\pm$   1.34 \\
       973 &      6.912 &      7.424 &     227.09 $\pm$  87.91 &      -1.14 $\pm$   0.25 &      -1.91 $\pm$   0.12 &       2.87 $\pm$   1.52 \\
       973 &      7.424 &      7.936 &     160.46 $\pm$  41.02 &      -0.88 $\pm$   0.35 &      -2.04 $\pm$   0.13 &        2.1 $\pm$   0.98 \\
       973 &      7.936 &      9.024 &     185.24 $\pm$  24.39 &      -1.09 $\pm$   0.16 &      -2.48 $\pm$    0.3 &       1.28 $\pm$   0.27 \\
       973 &      9.024 &      10.24 &      134.8 $\pm$     20 &      -0.73 $\pm$   0.32 &      -2.25 $\pm$   0.14 &       1.03 $\pm$   0.36 \\
       973 &      10.24 &     12.416 &      176.9 $\pm$  41.82 &      -1.57 $\pm$   0.15 &      -2.49 $\pm$   0.65 &       0.73 $\pm$   0.23 \\
       973 &     12.416 &     16.704 &     118.23 $\pm$  14.67 &      -1.11 $\pm$   0.29 &      -2.61 $\pm$   0.38 &       0.31 $\pm$   0.09 \\
      1625 &      0.064 &       4.16 &     630.13 $\pm$ 277.92 &      -0.99 $\pm$   0.16 &      -2.02 $\pm$    0.5 &       0.82 $\pm$   0.41 \\
      1625 &       4.16 &       4.48 &     865.53 $\pm$ 173.67 &      -0.63 $\pm$   0.08 &       -2.1 $\pm$   0.17 &      17.32 $\pm$   3.97 \\
      1625 &       4.48 &      4.672 &     926.02 $\pm$ 156.35 &      -0.52 $\pm$   0.07 &      -2.11 $\pm$   0.16 &      41.34 $\pm$   7.75 \\
      1625 &      4.672 &        4.8 &    1281.55 $\pm$ 300.49 &      -0.63 $\pm$   0.07 &      -2.27 $\pm$    1.8 &      54.93 $\pm$   13.7 \\
      1625 &        4.8 &      4.928 &     663.63 $\pm$ 108.87 &      -0.42 $\pm$    0.1 &      -1.75 $\pm$   0.11 &      65.92 $\pm$  11.73 \\
      1625 &      4.928 &       5.12 &     892.01 $\pm$ 128.88 &      -0.67 $\pm$   0.06 &      -1.98 $\pm$   0.19 &      57.54 $\pm$   8.97 \\
      1625 &       5.12 &      5.248 &     548.89 $\pm$  79.89 &      -0.58 $\pm$   0.09 &      -2.02 $\pm$   0.16 &      36.32 $\pm$   5.68 \\
      1625 &      5.248 &       5.44 &     625.82 $\pm$  85.39 &      -0.83 $\pm$   0.07 &      -2.37 $\pm$   0.33 &      21.29 $\pm$   3.12 \\
      1625 &       5.44 &      5.632 &     508.13 $\pm$  60.76 &      -0.75 $\pm$   0.07 &      -2.44 $\pm$    0.3 &      16.95 $\pm$   2.19 \\
      1625 &      5.632 &      5.824 &     548.99 $\pm$  89.78 &      -0.92 $\pm$   0.08 &      -2.32 $\pm$   0.32 &      14.96 $\pm$   2.58 \\
      1625 &      5.824 &       6.08 &     386.59 $\pm$  72.37 &      -1.02 $\pm$    0.1 &      -2.08 $\pm$   0.16 &      10.95 $\pm$   2.15 \\
      1625 &       6.08 &        6.4 &     568.82 $\pm$ 142.06 &      -1.21 $\pm$   0.08 &      -2.25 $\pm$    0.4 &       7.95 $\pm$   2.06 \\
      1625 &        6.4 &       6.72 &     347.37 $\pm$  58.06 &      -0.86 $\pm$   0.12 &      -2.14 $\pm$   0.18 &       6.87 $\pm$   1.21 \\
      1625 &       6.72 &      7.104 &     310.33 $\pm$  62.73 &      -1.13 $\pm$   0.12 &      -2.33 $\pm$   0.33 &       3.76 $\pm$   0.79 \\
      1625 &      7.104 &      7.552 &     196.74 $\pm$  44.88 &       -0.8 $\pm$   0.24 &      -1.93 $\pm$    0.1 &       4.15 $\pm$   0.99 \\
      1625 &      7.552 &       8.96 &     175.43 $\pm$  25.46 &      -0.92 $\pm$   0.19 &      -2.33 $\pm$   0.22 &       1.13 $\pm$   0.17 \\
      1733 &      -5.12 &      1.024 &      511.7 $\pm$    304 &       -1.2 $\pm$   0.18 &      -7.14 $\pm$      6 &       0.23 $\pm$   0.17 \\
      1733 &      1.024 &      3.072 &      693.1 $\pm$   96.9 &      -1.53 $\pm$    0.1 &       -1.9 $\pm$   0.17 &          2 $\pm$   0.35 \\
      1733 &      3.072 &       5.12 &       1042 $\pm$    280 &      -1.03 $\pm$   0.05 &      -1.98 $\pm$   0.39 &       4.02 $\pm$   1.23 \\
      1733 &       5.12 &      7.168 &      490.8 $\pm$    219 &      -1.23 $\pm$   0.11 &      -1.77 $\pm$   0.14 &       2.05 $\pm$   0.98 \\
      1733 &      7.168 &      9.216 &      320.8 $\pm$    114 &      -1.14 $\pm$   0.17 &      -2.13 $\pm$   0.44 &        0.7 $\pm$   0.29 \\
      1733 &     13.312 &     21.504 &      184.2 $\pm$   53.1 &      -0.78 $\pm$   0.37 &      -2.28 $\pm$   0.52 &       0.15 $\pm$   0.06 \\
      1883 &          0 &      0.576 &      478.5 $\pm$   60.9 &      -0.11 $\pm$   0.17 &      -3.13 $\pm$   1.15 &       2.11 $\pm$   0.39 \\
      1883 &      0.576 &      1.024 &      314.7 $\pm$   31.6 &      -0.15 $\pm$   0.16 &      -2.63 $\pm$   0.34 &          3 $\pm$   0.41 \\
      1883 &      1.024 &      1.472 &      235.2 $\pm$     23 &      -0.27 $\pm$   0.18 &      -2.27 $\pm$   0.23 &       3.37 $\pm$   0.45 \\
      1883 &      1.472 &       1.92 &      207.8 $\pm$   19.3 &      -0.36 $\pm$   0.18 &      -2.58 $\pm$   0.26 &       2.47 $\pm$   0.32 \\
      1883 &       1.92 &      2.432 &      180.1 $\pm$   16.7 &      -0.88 $\pm$   0.15 &      -3.02 $\pm$   0.65 &       1.72 $\pm$   0.23 \\
      1883 &      2.432 &      2.944 &      132.7 $\pm$   21.1 &      -0.33 $\pm$   0.32 &      -2.57 $\pm$   0.15 &       0.98 $\pm$    0.2 \\
      1883 &      2.944 &      3.584 &      135.5 $\pm$   19.6 &      -1.23 $\pm$   0.22 &      -2.63 $\pm$   0.45 &       1.19 $\pm$   0.24 \\
      1883 &      3.584 &      4.352 &      107.3 $\pm$   12.6 &      -1.27 $\pm$   0.26 &      -3.01 $\pm$   0.97 &       0.73 $\pm$   0.14 \\
      1956 &      0.064 &      1.728 &      114.2 $\pm$   33.7 &      -0.65 $\pm$   0.56 &      -1.97 $\pm$   0.11 &       0.78 $\pm$   0.27 \\
      1956 &      1.728 &      2.368 &      120.8 $\pm$   24.2 &       -0.6 $\pm$   0.33 &       -2.2 $\pm$   0.17 &          1 $\pm$   0.25 \\
      1956 &      2.368 &      2.944 &      135.6 $\pm$   20.1 &       0.37 $\pm$   0.96 &      -1.91 $\pm$   0.07 &       4.43 $\pm$   0.85 \\
      1956 &      2.944 &       3.52 &      138.8 $\pm$   32.2 &      -0.87 $\pm$   0.33 &      -2.08 $\pm$   0.13 &       1.61 $\pm$   0.45 \\
      1956 &       3.52 &       4.16 &      121.7 $\pm$   18.2 &      -0.59 $\pm$   0.25 &      -2.64 $\pm$   0.39 &       0.64 $\pm$   0.14 \\
      1956 &       4.16 &      4.864 &      108.9 $\pm$   26.5 &      -0.64 $\pm$   0.52 &      -2.08 $\pm$   0.12 &       1.14 $\pm$   0.34 \\
      1956 &      4.864 &      5.632 &      65.13 $\pm$   19.1 &       0.37 $\pm$   1.92 &      -2.08 $\pm$   0.09 &       0.78 $\pm$   0.28 \\
      1956 &      5.632 &      6.464 &      74.97 $\pm$   11.7 &       0.62 $\pm$    1.2 &      -2.38 $\pm$   0.15 &       0.48 $\pm$   0.11 \\
     2083 &          0 &      0.256 &     660.98 $\pm$ 100.06 &      -0.74 $\pm$   0.08 &      -5.14 $\pm$   6.72 &       6.15 $\pm$    1.1 \\
      2083 &      0.256 &      0.448 &     658.86 $\pm$   80.9 &      -0.58 $\pm$   0.07 &      -2.45 $\pm$   0.37 &      17.84 $\pm$   2.63 \\
      2083 &      0.704 &       0.96 &      541.2 $\pm$  21.95 &      -0.13 $\pm$   0.05 &      -3.27 $\pm$   0.37 &      36.71 $\pm$   2.01 \\
      2083 &       0.96 &      1.088 &     518.05 $\pm$  28.76 &      -0.14 $\pm$   0.06 &      -3.15 $\pm$   0.41 &      41.49 $\pm$   3.14 \\
      2083 &      1.088 &      1.216 &     473.24 $\pm$  20.47 &       -0.1 $\pm$   0.06 &      -4.18 $\pm$   1.13 &      37.12 $\pm$   2.29 \\
      2083 &      1.216 &      1.344 &     397.75 $\pm$  18.48 &      -0.06 $\pm$   0.07 &       -3.4 $\pm$   0.43 &      31.52 $\pm$   2.22 \\
      2083 &      1.344 &      1.536 &     321.28 $\pm$  12.52 &      -0.13 $\pm$   0.06 &      -3.29 $\pm$   0.31 &      22.55 $\pm$   1.45 \\
      2083 &      1.536 &      1.664 &     282.47 $\pm$  16.13 &      -0.36 $\pm$   0.09 &      -3.07 $\pm$   0.33 &      16.72 $\pm$   1.58 \\
      2083 &      1.792 &       1.92 &     233.34 $\pm$  14.64 &      -0.49 $\pm$   0.11 &      -3.07 $\pm$   0.38 &      10.68 $\pm$    1.2 \\
      2083 &       1.92 &      2.112 &     193.38 $\pm$    9.8 &      -0.23 $\pm$   0.12 &      -2.74 $\pm$   0.17 &       9.77 $\pm$   1.07 \\
      2083 &      2.112 &       2.24 &     193.46 $\pm$  11.33 &      -0.19 $\pm$   0.14 &      -2.88 $\pm$   0.25 &       8.75 $\pm$   1.13 \\
      2083 &       2.24 &      2.368 &      182.1 $\pm$  11.48 &      -0.36 $\pm$   0.15 &      -2.85 $\pm$   0.25 &       8.14 $\pm$   1.13 \\
      2083 &      2.368 &       2.56 &     171.41 $\pm$   9.84 &      -0.32 $\pm$   0.14 &      -2.64 $\pm$   0.15 &       8.05 $\pm$   1.07 \\
      2083 &       2.56 &      2.752 &     159.63 $\pm$   8.85 &       -0.5 $\pm$   0.14 &      -2.85 $\pm$   0.22 &       6.44 $\pm$   0.84 \\
      2083 &      2.752 &      2.944 &      160.6 $\pm$  10.07 &       -0.8 $\pm$   0.13 &      -2.93 $\pm$   0.31 &       5.92 $\pm$   0.78 \\
      2083 &      2.944 &      3.136 &     148.76 $\pm$   8.08 &      -0.91 $\pm$   0.12 &      -3.41 $\pm$   0.69 &        4.9 $\pm$   0.58 \\
      2083 &      3.136 &      3.328 &      127.1 $\pm$   5.93 &       -0.7 $\pm$   0.15 &      -3.36 $\pm$   0.49 &        4.1 $\pm$   0.55 \\
      2083 &      3.328 &      3.584 &     117.85 $\pm$   6.14 &      -0.89 $\pm$   0.16 &      -2.97 $\pm$   0.28 &       3.95 $\pm$   0.57 \\
      2083 &      3.584 &       3.84 &      106.5 $\pm$   4.62 &      -0.62 $\pm$   0.19 &       -3.1 $\pm$   0.27 &       3.33 $\pm$   0.53 \\
      2083 &       3.84 &      4.096 &     102.21 $\pm$   4.95 &      -1.06 $\pm$   0.17 &      -3.18 $\pm$   0.43 &       3.34 $\pm$   0.49 \\
      2083 &      4.096 &      4.352 &       87.5 $\pm$    4.3 &      -1.03 $\pm$   0.22 &      -3.04 $\pm$    0.3 &       3.01 $\pm$   0.56 \\
      2083 &      4.352 &      4.672 &      72.33 $\pm$   4.27 &      -1.18 $\pm$   0.27 &      -2.78 $\pm$   0.17 &       3.08 $\pm$   0.74 \\
      2083 &      4.672 &      4.992 &       73.2 $\pm$   2.92 &      -0.27 $\pm$   0.41 &      -3.03 $\pm$   0.17 &       1.97 $\pm$   0.66 \\
      2083 &      4.992 &      5.376 &      68.51 $\pm$   3.95 &      -1.06 $\pm$   0.28 &      -2.99 $\pm$   0.22 &       2.18 $\pm$   0.55 \\
      2083 &      5.376 &       5.76 &       63.7 $\pm$   4.52 &       -1.1 $\pm$    0.3 &       -3.1 $\pm$   0.26 &       2.03 $\pm$   0.56 \\
      2083 &       5.76 &      6.144 &      68.02 $\pm$   3.57 &      -0.92 $\pm$   0.35 &       -2.9 $\pm$   0.18 &       2.08 $\pm$   0.62 \\
      2193 &          0 &       3.84 &      552.6 $\pm$   96.4 &       0.18 $\pm$   0.19 &      -3.75 $\pm$   0.46 &       0.51 $\pm$   0.12 \\
      2193 &       3.84 &      5.952 &      459.6 $\pm$     40 &        0.1 $\pm$   0.24 &      -3.12 $\pm$   0.74 &        0.7 $\pm$   0.09 \\
      2193 &      5.952 &       7.36 &      405.9 $\pm$   32.7 &       0.87 $\pm$   0.24 &      -4.08 $\pm$   2.31 &        0.8 $\pm$   0.11 \\
      2193 &       7.36 &      8.704 &      387.8 $\pm$   36.5 &       0.67 $\pm$   0.24 &      -3.16 $\pm$   0.85 &       0.89 $\pm$   0.13 \\
      2193 &      8.704 &      9.984 &      335.6 $\pm$   31.4 &       1.06 $\pm$   0.31 &      -2.59 $\pm$   0.36 &       1.05 $\pm$   0.15 \\
      2193 &      9.984 &     11.264 &      362.6 $\pm$   41.2 &       0.54 $\pm$   0.26 &      -2.51 $\pm$   0.39 &       1.07 $\pm$   0.17 \\
      2193 &     11.264 &     12.608 &      360.8 $\pm$   27.3 &        0.5 $\pm$   0.21 &      -3.19 $\pm$    0.3 &       0.78 $\pm$   0.11 \\
      2193 &     12.608 &     13.888 &      320.2 $\pm$   31.1 &       0.76 $\pm$   0.28 &      -2.56 $\pm$   0.36 &          1 $\pm$   0.14 \\
      2193 &     13.888 &     15.232 &      261.7 $\pm$   19.5 &       1.26 $\pm$   0.33 &      -2.88 $\pm$   0.45 &        0.7 $\pm$   0.09 \\
      2193 &     15.232 &      16.64 &      257.3 $\pm$     18 &       1.31 $\pm$   0.33 &       -3.2 $\pm$   0.66 &       0.58 $\pm$   0.08 \\
      2193 &      16.64 &     18.048 &      234.6 $\pm$   22.8 &       1.56 $\pm$   0.42 &      -3.17 $\pm$   0.21 &       0.46 $\pm$   0.07 \\
      2193 &     18.048 &     19.456 &      259.9 $\pm$   21.4 &       0.76 $\pm$   0.28 &      -3.04 $\pm$   0.65 &       0.58 $\pm$   0.08 \\
      2193 &     19.456 &     20.864 &      261.5 $\pm$   16.4 &       0.79 $\pm$   0.25 &      -6.04 $\pm$    3.5 &       0.45 $\pm$   0.06 \\
      2193 &     20.864 &     22.912 &      197.5 $\pm$   12.9 &       1.59 $\pm$    0.4 &      -2.81 $\pm$   0.34 &       0.45 $\pm$   0.06 \\
      2193 &     22.912 &      24.32 &      198.8 $\pm$   15.9 &       0.97 $\pm$   0.38 &      -3.28 $\pm$   0.86 &       0.35 $\pm$   0.06 \\
      2193 &      24.32 &     25.728 &      184.1 $\pm$   14.4 &       1.05 $\pm$    0.4 &      -3.05 $\pm$   0.58 &       0.38 $\pm$   0.06 \\
      2193 &     25.728 &     27.136 &      178.9 $\pm$   14.6 &       0.97 $\pm$   0.41 &      -3.07 $\pm$   0.62 &       0.34 $\pm$   0.06 \\
      2193 &     27.136 &     30.016 &      166.2 $\pm$   10.3 &       1.62 $\pm$   0.42 &      -3.06 $\pm$   0.27 &        0.3 $\pm$   0.04 \\
      2193 &     33.024 &      37.76 &      147.6 $\pm$   8.34 &       1.56 $\pm$   0.38 &      -3.13 $\pm$   0.43 &       0.16 $\pm$   0.02 \\
      2193 &      37.76 &      40.96 &      132.4 $\pm$   7.57 &       1.32 $\pm$   0.47 &      -3.86 $\pm$    1.1 &       0.14 $\pm$   0.02 \\
      2387 &          0 &      2.432 &     288.77 $\pm$  40.35 &          0 $\pm$   0.22 &       -1.9 $\pm$   0.13 &       1.64 $\pm$   0.39 \\
      2387 &      2.432 &      3.392 &      146.2 $\pm$  19.99 &       0.78 $\pm$   0.53 &      -1.78 $\pm$   0.06 &       2.94 $\pm$   1.34 \\
      2387 &      3.392 &      4.288 &     211.88 $\pm$  19.35 &      -0.06 $\pm$    0.2 &      -2.32 $\pm$   0.18 &        1.9 $\pm$   0.36 \\
      2387 &      4.288 &      5.184 &     179.97 $\pm$  13.48 &       0.16 $\pm$   0.21 &      -2.39 $\pm$   0.15 &       1.81 $\pm$   0.34 \\
      2387 &      5.184 &      6.016 &     167.83 $\pm$  12.41 &        0.2 $\pm$   0.23 &      -2.43 $\pm$   0.16 &       1.71 $\pm$   0.34 \\
      2387 &      6.016 &      6.848 &     199.05 $\pm$  14.84 &       -0.2 $\pm$   0.17 &      -2.69 $\pm$   0.29 &       1.58 $\pm$   0.25 \\
      2387 &      6.848 &      7.744 &      169.8 $\pm$  12.81 &      -0.13 $\pm$    0.2 &      -2.48 $\pm$   0.18 &       1.66 $\pm$    0.3 \\
      2387 &      7.744 &       8.64 &     170.23 $\pm$  12.91 &      -0.13 $\pm$    0.2 &      -2.52 $\pm$    0.2 &       1.53 $\pm$   0.28 \\
      2387 &       8.64 &      9.536 &     159.84 $\pm$  13.21 &       -0.2 $\pm$   0.22 &      -2.48 $\pm$   0.19 &       1.44 $\pm$    0.3 \\
      2387 &      9.536 &     10.432 &     153.53 $\pm$  12.32 &       0.04 $\pm$   0.25 &      -2.42 $\pm$   0.16 &       1.47 $\pm$   0.33 \\
      2387 &     10.432 &     11.328 &     161.48 $\pm$  15.23 &      -0.41 $\pm$   0.22 &      -2.49 $\pm$   0.22 &       1.32 $\pm$   0.29 \\
      2387 &     11.328 &     12.224 &     145.46 $\pm$   11.8 &      -0.22 $\pm$   0.25 &      -2.61 $\pm$   0.24 &       1.11 $\pm$   0.25 \\
      2387 &     12.224 &     13.248 &     134.81 $\pm$   11.2 &      -0.47 $\pm$   0.24 &      -2.63 $\pm$   0.25 &       1.01 $\pm$   0.23 \\
      2387 &     13.248 &     14.272 &     107.57 $\pm$   9.32 &        0.4 $\pm$   0.49 &      -2.47 $\pm$   0.15 &       0.85 $\pm$   0.35 \\
      2387 &     14.272 &     15.808 &     123.71 $\pm$  12.88 &      -0.62 $\pm$   0.28 &      -2.41 $\pm$   0.17 &       0.89 $\pm$   0.25 \\
      2387 &     15.808 &     18.048 &      98.49 $\pm$   4.85 &       0.47 $\pm$   0.37 &      -2.87 $\pm$   0.18 &        0.5 $\pm$   0.15 \\
      2387 &     18.048 &      20.48 &      88.73 $\pm$   5.13 &       0.27 $\pm$   0.47 &      -2.86 $\pm$   0.21 &       0.39 $\pm$   0.15 \\
      2387 &      20.48 &     25.856 &      75.48 $\pm$   4.58 &       0.96 $\pm$   0.76 &      -2.73 $\pm$   0.14 &       0.26 $\pm$   0.16 \\
      2387 &     25.856 &     38.208 &       68.4 $\pm$   2.88 &       1.36 $\pm$   0.89 &      -2.51 $\pm$   0.35 &       0.16 $\pm$   0.11 \\
      3003 &          0 &        6.4 &      759.9 $\pm$ 345.34 &      -0.89 $\pm$   0.13 &      -1.57 $\pm$   0.13 &       1.48 $\pm$   0.75 \\
      3003 &        6.4 &      7.808 &     542.53 $\pm$  152.1 &      -0.85 $\pm$   0.12 &      -2.73 $\pm$   0.13 &        1.6 $\pm$   0.53 \\
      3003 &      7.808 &      8.704 &     640.64 $\pm$ 175.73 &      -0.92 $\pm$    0.1 &      -2.98 $\pm$   0.28 &        2.1 $\pm$   0.66 \\
      3003 &      8.704 &        9.6 &     997.36 $\pm$ 379.48 &      -1.11 $\pm$   0.08 &       -2.1 $\pm$   0.65 &       3.48 $\pm$   1.42 \\
      3003 &        9.6 &     10.496 &    1126.81 $\pm$ 695.77 &      -1.28 $\pm$   0.08 &      -2.61 $\pm$    5.2 &       2.68 $\pm$   1.74 \\
      3003 &     10.496 &     11.392 &     465.62 $\pm$ 215.46 &      -1.13 $\pm$   0.14 &      -1.71 $\pm$   0.11 &       3.51 $\pm$   1.87 \\
      3003 &     11.392 &     12.288 &     261.84 $\pm$  52.21 &      -0.83 $\pm$   0.17 &         -2 $\pm$   0.15 &       2.08 $\pm$   0.59 \\
      3003 &     12.288 &     13.184 &      591.4 $\pm$ 250.03 &      -1.25 $\pm$    0.1 &      -2.02 $\pm$   0.36 &       2.34 $\pm$    1.1 \\
      3003 &     13.184 &     14.656 &     353.46 $\pm$   84.7 &       -1.2 $\pm$   0.11 &      -2.25 $\pm$   0.38 &       1.35 $\pm$   0.39 \\
      3003 &     14.656 &     16.256 &      314.7 $\pm$  80.18 &      -1.14 $\pm$   0.13 &      -2.17 $\pm$   0.31 &       1.13 $\pm$   0.36 \\
      3003 &     16.256 &     19.136 &      317.1 $\pm$  55.69 &      -1.22 $\pm$   0.09 &      -3.25 $\pm$   3.04 &       0.63 $\pm$   0.14 \\
      3003 &     19.136 &     24.512 &     321.97 $\pm$  94.62 &      -1.43 $\pm$   0.11 &      -2.96 $\pm$    3.3 &       0.39 $\pm$   0.14 \\
      3658 &      0.026 &       0.32 &      856.5 $\pm$    131 &       -0.3 $\pm$   0.11 &      -3.25 $\pm$   2.63 &       6.72 $\pm$   1.34 \\
      3658 &       0.32 &      0.512 &      531.2 $\pm$   97.8 &      -0.32 $\pm$   0.14 &      -1.88 $\pm$   0.17 &      13.68 $\pm$    3.1 \\
      3658 &      0.512 &      0.704 &      408.1 $\pm$   53.1 &      -0.02 $\pm$   0.15 &       -1.9 $\pm$   0.12 &      15.28 $\pm$   2.49 \\
      3658 &      0.704 &      0.832 &      373.7 $\pm$   52.4 &      -0.15 $\pm$   0.16 &      -2.03 $\pm$   0.16 &       14.2 $\pm$    2.5 \\
      3658 &      0.832 &      1.024 &      349.3 $\pm$   32.3 &      -0.25 $\pm$   0.11 &      -2.49 $\pm$   0.25 &        9.1 $\pm$   1.13 \\
      3658 &      1.024 &      1.216 &      240.5 $\pm$   27.3 &      -0.25 $\pm$   0.17 &      -2.22 $\pm$   0.17 &       6.22 $\pm$   0.92 \\
      3658 &      1.216 &      1.472 &      178.1 $\pm$   24.9 &      -0.48 $\pm$   0.23 &      -2.22 $\pm$   0.18 &       3.23 $\pm$   0.59 \\
      3658 &      1.472 &       1.92 &      156.7 $\pm$   35.6 &      -0.82 $\pm$   0.29 &      -2.06 $\pm$   0.14 &       2.08 $\pm$   0.57 \\
      3658 &       1.92 &       2.56 &      124.6 $\pm$   31.8 &      -0.73 $\pm$   0.46 &      -2.11 $\pm$   0.16 &       1.07 $\pm$   0.34 \\
      3765 &     62.208 &     62.912 &     126.96 $\pm$  27.61 &      -0.95 $\pm$   0.36 &      -2.44 $\pm$   0.13 &       1.21 $\pm$   0.55 \\
      3765 &     62.912 &     63.552 &     169.45 $\pm$  15.98 &      -1.02 $\pm$   0.15 &      -3.61 $\pm$   2.32 &       1.13 $\pm$    0.2 \\
      3765 &     63.552 &         64 &      92.31 $\pm$  15.69 &       0.68 $\pm$   0.93 &      -3.43 $\pm$   0.06 &       0.94 $\pm$   0.82 \\
      3765 &         64 &      64.32 &     228.57 $\pm$  42.74 &      -1.06 $\pm$   0.16 &      -2.26 $\pm$   0.23 &       3.55 $\pm$   0.96 \\
      3765 &      64.32 &     64.576 &     255.67 $\pm$  27.25 &      -0.93 $\pm$   0.12 &       -3.8 $\pm$   2.98 &       2.91 $\pm$   0.47 \\
      3765 &     64.576 &     64.832 &     239.14 $\pm$  38.74 &      -0.92 $\pm$   0.15 &      -2.27 $\pm$   0.21 &       4.64 $\pm$   1.12 \\
      3765 &     64.832 &     65.024 &     367.07 $\pm$  54.23 &      -0.92 $\pm$   0.11 &      -2.83 $\pm$   0.76 &       5.51 $\pm$   1.06 \\
      3765 &     65.024 &     65.216 &     309.53 $\pm$  42.03 &      -0.73 $\pm$   0.12 &      -2.21 $\pm$   0.17 &       9.58 $\pm$   1.84 \\
      3765 &     65.216 &     65.408 &     374.47 $\pm$  32.26 &      -0.84 $\pm$   0.07 &      -4.43 $\pm$   5.87 &       7.25 $\pm$   0.84 \\
      3765 &     65.408 &     65.536 &     349.29 $\pm$  40.22 &      -0.71 $\pm$   0.11 &      -2.81 $\pm$   0.51 &       9.35 $\pm$   1.49 \\
      3765 &     65.536 &     65.792 &     358.48 $\pm$  28.66 &      -0.71 $\pm$   0.07 &      -2.78 $\pm$   0.33 &      10.39 $\pm$   1.14 \\
      3765 &     65.792 &      65.92 &     416.84 $\pm$  42.29 &      -0.69 $\pm$   0.08 &      -2.87 $\pm$   0.52 &      14.45 $\pm$   1.94 \\
      3765 &      65.92 &     66.048 &     402.24 $\pm$  39.42 &      -0.55 $\pm$   0.09 &      -2.54 $\pm$   0.27 &      18.63 $\pm$   2.48 \\
      3765 &     66.048 &     66.176 &      353.8 $\pm$  34.24 &      -0.56 $\pm$   0.09 &      -2.46 $\pm$   0.22 &      18.12 $\pm$   2.48 \\
      3765 &     66.176 &     66.368 &     460.24 $\pm$  44.61 &      -0.74 $\pm$   0.07 &      -2.52 $\pm$   0.27 &      19.03 $\pm$   2.34 \\
      3765 &     66.368 &     66.496 &     415.71 $\pm$  49.56 &      -0.75 $\pm$   0.09 &      -2.51 $\pm$   0.31 &      16.85 $\pm$   2.61 \\
      3765 &     66.496 &     66.624 &     259.13 $\pm$  28.85 &       -0.4 $\pm$   0.15 &      -2.14 $\pm$   0.12 &      15.52 $\pm$   2.84 \\
      3765 &     66.624 &     66.816 &     306.45 $\pm$  31.06 &      -0.69 $\pm$    0.1 &      -2.37 $\pm$   0.18 &      11.76 $\pm$   1.72 \\
      3765 &     66.816 &     67.072 &     280.38 $\pm$  26.84 &      -0.76 $\pm$   0.09 &      -2.37 $\pm$   0.16 &        9.6 $\pm$   1.35 \\
      3765 &     67.072 &     67.328 &      310.7 $\pm$  25.73 &      -0.89 $\pm$   0.08 &      -3.42 $\pm$   1.11 &       6.33 $\pm$   0.73 \\
      3765 &     67.328 &     67.648 &     294.92 $\pm$  32.75 &      -0.97 $\pm$   0.09 &      -2.55 $\pm$   0.29 &       6.15 $\pm$   0.94 \\
      3765 &     67.648 &     68.096 &     206.78 $\pm$  23.17 &      -0.89 $\pm$   0.12 &      -2.24 $\pm$   0.12 &       5.04 $\pm$    0.9 \\
      3765 &     68.096 &     68.672 &     215.82 $\pm$   18.2 &      -0.97 $\pm$    0.1 &       -2.9 $\pm$   0.48 &       2.71 $\pm$   0.36 \\
      3765 &     68.672 &     69.568 &     171.33 $\pm$  19.19 &      -1.05 $\pm$   0.14 &       -2.4 $\pm$   0.19 &       2.05 $\pm$   0.39 \\
      3765 &     69.568 &     70.912 &     146.77 $\pm$  13.54 &      -1.14 $\pm$   0.14 &      -2.72 $\pm$   0.38 &       1.13 $\pm$   0.19 \\
      3765 &     70.912 &      73.28 &     109.87 $\pm$   9.59 &         -1 $\pm$   0.23 &       -2.6 $\pm$   0.23 &       0.67 $\pm$   0.15 \\
      5478 &      0.029 &      0.704 &      418.7 $\pm$    193 &      -0.29 $\pm$   0.38 &      -3.51 $\pm$   0.15 &        0.9 $\pm$   0.47 \\
      5478 &      0.704 &      1.728 &      285.9 $\pm$   19.4 &       0.05 $\pm$   0.15 &     -10.94 $\pm$     80 &       1.02 $\pm$   0.12 \\
      5478 &      1.728 &      3.136 &      193.9 $\pm$   13.6 &       0.51 $\pm$   0.22 &      -2.48 $\pm$   0.21 &        1.3 $\pm$   0.15 \\
      5478 &      3.136 &      4.224 &      175.5 $\pm$   12.4 &       0.25 $\pm$   0.24 &      -3.01 $\pm$   0.72 &       0.79 $\pm$   0.11 \\
      5478 &      4.224 &      4.864 &      113.4 $\pm$   17.7 &       0.86 $\pm$   0.85 &       -2.4 $\pm$   0.18 &       0.72 $\pm$   0.17 \\
      5478 &      4.864 &      6.336 &      112.9 $\pm$   10.1 &        1.5 $\pm$   0.69 &      -2.45 $\pm$   0.18 &       0.57 $\pm$   0.09 \\
      5478 &      6.336 &      9.408 &      95.91 $\pm$   9.67 &       1.62 $\pm$   0.94 &      -2.47 $\pm$   0.19 &        0.3 $\pm$   0.05 \\
      5478 &      9.408 &     18.624 &      65.75 $\pm$   10.3 &       0.04 $\pm$   1.36 &       -2.6 $\pm$   0.28 &       0.11 $\pm$   0.03 \\
      5523 &      0.033 &      0.512 &      73.84 $\pm$   19.4 &       1.61 $\pm$   2.36 &      -4.92 $\pm$   0.09 &       0.39 $\pm$   0.13 \\
      5523 &      0.512 &       0.96 &      165.1 $\pm$   38.3 &      -0.55 $\pm$   0.36 &      -3.54 $\pm$   0.17 &       1.11 $\pm$   0.32 \\
      5523 &       0.96 &      1.408 &      218.6 $\pm$   52.2 &      -0.95 $\pm$   0.22 &      -3.25 $\pm$   0.35 &       1.47 $\pm$   0.44 \\
      5523 &      1.408 &      1.856 &      121.8 $\pm$     21 &       0.09 $\pm$   0.79 &      -2.08 $\pm$   0.13 &        2.6 $\pm$   0.61 \\
      5523 &      1.856 &      2.368 &      101.2 $\pm$   22.8 &      -0.02 $\pm$   0.83 &       -2.1 $\pm$   0.14 &        1.5 $\pm$   0.44 \\
      5523 &      2.368 &      3.008 &      94.75 $\pm$   20.6 &      -0.09 $\pm$   0.86 &      -2.19 $\pm$   0.17 &       1.06 $\pm$   0.31 \\
      5523 &      3.008 &        4.8 &      76.69 $\pm$   12.9 &      -0.11 $\pm$   0.98 &      -2.21 $\pm$    0.2 &       0.59 $\pm$   0.15 \\
      5523 &        4.8 &      9.472 &      64.26 $\pm$   8.14 &       0.15 $\pm$   1.58 &      -2.12 $\pm$   1.68 &       0.21 $\pm$   0.06 \\
      5601 &      5.056 &      6.016 &      189.6 $\pm$   57.8 &      -1.26 $\pm$   0.21 &      -3.19 $\pm$    0.4 &       0.55 $\pm$   0.22 \\
      5601 &      6.016 &      6.592 &      318.4 $\pm$    129 &      -0.95 $\pm$   0.17 &       -2.4 $\pm$   0.94 &       1.25 $\pm$    0.6 \\
      5601 &      6.592 &      6.976 &      465.2 $\pm$   92.5 &      -0.66 $\pm$   0.16 &      -2.67 $\pm$   0.83 &       3.08 $\pm$   0.79 \\
      5601 &      6.976 &       7.36 &      403.1 $\pm$     99 &      -0.61 $\pm$   0.19 &      -1.98 $\pm$    0.2 &       5.04 $\pm$   1.44 \\
      5601 &       7.36 &      7.744 &      713.8 $\pm$    224 &      -0.91 $\pm$   0.13 &      -2.23 $\pm$   0.61 &       5.04 $\pm$   1.88 \\
      5601 &      7.744 &      8.128 &      432.3 $\pm$   84.6 &      -0.99 $\pm$   0.12 &      -6.74 $\pm$      7 &        2.6 $\pm$   0.68 \\
      5601 &      8.128 &      8.512 &      271.1 $\pm$   71.3 &      -0.76 $\pm$   0.24 &      -2.05 $\pm$   0.18 &       3.54 $\pm$   1.07 \\
      5601 &      8.512 &       8.96 &      193.1 $\pm$   42.8 &      -0.65 $\pm$   0.31 &      -2.07 $\pm$   0.16 &        2.8 $\pm$   0.74 \\
      5601 &       8.96 &      9.472 &      172.5 $\pm$   49.3 &      -0.98 $\pm$   0.32 &      -2.08 $\pm$   0.16 &       2.32 $\pm$   0.77 \\
      5601 &      9.472 &     10.048 &      147.1 $\pm$   39.8 &      -1.02 $\pm$   0.37 &      -2.18 $\pm$    0.2 &       2.32 $\pm$   0.77 \\
      5601 &     10.048 &     10.688 &      212.8 $\pm$    154 &      -1.42 $\pm$   0.41 &      -1.87 $\pm$   0.14 &       1.67 $\pm$   0.54 \\
      5601 &     10.688 &      11.52 &      106.4 $\pm$     51 &      -1.45 $\pm$   1.05 &      -1.94 $\pm$   0.11 &       1.67 $\pm$   0.54 \\
      5601 &      11.52 &     13.888 &      53.59 $\pm$   35.5 &      -1.48 $\pm$   3.28 &      -2.03 $\pm$   0.07 &       2.18 $\pm$    1.7 \\
      6397 &      0.092 &      1.792 &      294.6 $\pm$   29.9 &       -0.4 $\pm$   0.11 &      -2.17 $\pm$   0.18 &       2.24 $\pm$   0.29 \\
      6397 &      1.792 &      2.368 &      246.4 $\pm$   26.3 &      -0.28 $\pm$   0.15 &      -2.17 $\pm$   0.17 &       3.52 $\pm$    0.5 \\
      6397 &      2.368 &       2.88 &      265.8 $\pm$   26.4 &      -0.53 $\pm$   0.12 &      -2.59 $\pm$   0.38 &       2.92 $\pm$    0.4 \\
      6397 &       2.88 &      3.392 &      200.1 $\pm$   22.8 &      -0.24 $\pm$   0.19 &      -2.03 $\pm$   0.11 &       4.17 $\pm$   0.62 \\
      6397 &      3.392 &      3.904 &      204.5 $\pm$     24 &      -0.26 $\pm$   0.19 &      -2.01 $\pm$   0.11 &        4.3 $\pm$   0.65 \\
      6397 &      3.904 &      4.416 &      204.4 $\pm$   19.2 &      -0.28 $\pm$   0.16 &      -2.34 $\pm$    0.2 &       2.87 $\pm$   0.38 \\
      6397 &      4.416 &      4.992 &      156.5 $\pm$   17.9 &       -0.1 $\pm$   0.25 &      -2.04 $\pm$    0.1 &       3.18 $\pm$   0.48 \\
      6397 &      4.992 &      5.632 &      212.7 $\pm$   21.2 &      -0.56 $\pm$   0.14 &      -2.54 $\pm$   0.33 &       2.07 $\pm$    0.3 \\
      6397 &      5.632 &      6.272 &      168.9 $\pm$   21.2 &      -0.49 $\pm$    0.2 &      -2.15 $\pm$   0.14 &       2.45 $\pm$   0.41 \\
      6397 &      6.272 &      6.976 &      186.4 $\pm$   20.4 &      -0.74 $\pm$   0.15 &      -2.48 $\pm$    0.3 &       1.82 $\pm$   0.28 \\
      6397 &      6.976 &      7.744 &      169.7 $\pm$   15.3 &      -0.54 $\pm$   0.16 &      -2.64 $\pm$   0.35 &       1.45 $\pm$    0.2 \\
      6397 &      7.744 &      8.512 &        178 $\pm$   16.2 &      -0.72 $\pm$   0.14 &       -3.1 $\pm$   0.94 &       1.17 $\pm$   0.18 \\
      6397 &      8.512 &      9.408 &      145.7 $\pm$   13.7 &      -0.47 $\pm$    0.2 &      -2.52 $\pm$   0.26 &        1.2 $\pm$   0.18 \\
      6397 &      9.408 &     10.368 &      92.93 $\pm$   11.8 &       0.48 $\pm$   0.62 &       -2.1 $\pm$   0.09 &       1.33 $\pm$   0.23 \\
      6397 &     10.368 &     12.864 &      131.8 $\pm$   7.56 &      -0.64 $\pm$   0.15 &      -3.52 $\pm$   1.21 &       0.53 $\pm$   0.06 \\
      6397 &     12.864 &     18.048 &      93.83 $\pm$   11.9 &      -0.52 $\pm$   0.37 &      -2.21 $\pm$    0.1 &       0.49 $\pm$   0.08 \\
      6397 &     18.048 &     38.464 &      77.83 $\pm$     13 &      -0.24 $\pm$   0.71 &      -2.17 $\pm$    0.1 &       0.19 $\pm$   0.04 \\
      6504 &       0.09 &        1.6 &      302.6 $\pm$   43.5 &       0.66 $\pm$   0.39 &       -2.5 $\pm$   0.59 &       0.89 $\pm$   0.19 \\
      6504 &        1.6 &      2.688 &      273.7 $\pm$     34 &       0.48 $\pm$   0.29 &       -2.4 $\pm$   0.33 &        1.4 $\pm$   0.25 \\
      6504 &      2.688 &      3.776 &      222.4 $\pm$   21.8 &       1.53 $\pm$   0.33 &      -2.08 $\pm$   0.34 &       1.53 $\pm$   0.25 \\
      6504 &      3.776 &      4.928 &      187.1 $\pm$   15.4 &       1.81 $\pm$   0.78 &      -2.65 $\pm$   0.17 &       1.85 $\pm$   0.27 \\
      6504 &      4.928 &      6.208 &        155 $\pm$   14.9 &       1.05 $\pm$    0.5 &      -2.48 $\pm$   0.28 &       0.75 $\pm$   0.13 \\
      6504 &      6.208 &       7.68 &      147.8 $\pm$   12.1 &       1.14 $\pm$    0.5 &      -2.96 $\pm$   0.53 &       0.45 $\pm$   0.08 \\
      6504 &     11.264 &     21.376 &       87.2 $\pm$   3.97 &       0.43 $\pm$   0.47 &      -5.94 $\pm$   13.8 &       0.09 $\pm$   0.01 \\
      6504 &     21.376 &     31.104 &      44.51 $\pm$   19.6 &       1.29 $\pm$   14.6 &      -2.93 $\pm$      1 &       0.02 $\pm$   0.02 \\
      6630 &      0.042 &      0.512 &      443.1 $\pm$  33.28 &       0.85 $\pm$   0.19 &      -2.62 $\pm$   0.27 &       6.28 $\pm$   0.75 \\
      6630 &      0.512 &      0.896 &     412.24 $\pm$  18.16 &       0.81 $\pm$   0.13 &       -4.3 $\pm$   1.28 &        6.4 $\pm$   0.48 \\
      6630 &      0.896 &      1.216 &     323.82 $\pm$  17.42 &       1.06 $\pm$   0.18 &      -2.71 $\pm$    0.2 &       8.14 $\pm$   0.88 \\
      6630 &      1.216 &      1.472 &     347.15 $\pm$  13.32 &       0.78 $\pm$   0.13 &      -5.96 $\pm$   6.63 &        6.7 $\pm$   0.49 \\
      6630 &      1.472 &      1.728 &     344.24 $\pm$  17.54 &       0.43 $\pm$   0.12 &      -3.43 $\pm$   0.52 &       8.14 $\pm$   0.73 \\
      6630 &      1.728 &      1.984 &     280.55 $\pm$  13.14 &       0.62 $\pm$   0.14 &      -2.99 $\pm$   0.33 &        7.9 $\pm$   0.77 \\
      6630 &      1.984 &       2.24 &     248.16 $\pm$  11.44 &        0.6 $\pm$   0.15 &       -3.1 $\pm$   0.29 &       6.84 $\pm$   0.72 \\
      6630 &       2.24 &      2.496 &     210.98 $\pm$   9.14 &       0.54 $\pm$   0.16 &      -3.16 $\pm$   0.29 &       5.88 $\pm$   0.66 \\
      6630 &      2.496 &      2.752 &     188.49 $\pm$   9.56 &       0.11 $\pm$   0.16 &      -3.13 $\pm$   0.33 &       4.65 $\pm$    0.6 \\
      6630 &      2.752 &      3.072 &     157.58 $\pm$   5.41 &       0.36 $\pm$   0.17 &         -4 $\pm$   0.73 &       3.06 $\pm$   0.36 \\
      6630 &      3.072 &      3.456 &     133.01 $\pm$   4.28 &       0.34 $\pm$   0.19 &      -3.93 $\pm$    0.6 &       2.37 $\pm$   0.32 \\
      6630 &      3.456 &      3.968 &     111.32 $\pm$   3.84 &       0.62 $\pm$   0.26 &      -3.35 $\pm$   0.25 &       1.85 $\pm$   0.36 \\
      6630 &      3.968 &      4.672 &      88.63 $\pm$   2.66 &      -0.05 $\pm$   0.26 &      -3.58 $\pm$   0.34 &       1.38 $\pm$   0.27 \\
      6630 &      4.672 &      5.696 &       65.6 $\pm$   2.28 &      -0.38 $\pm$   0.28 &       -4.2 $\pm$   0.51 &       1.04 $\pm$   0.24 \\
      6630 &      5.696 &     10.816 &      31.52 $\pm$   6.97 &      -0.14 $\pm$   1.13 &         -5 $\pm$   0.19 &       0.32 $\pm$   0.39 \\
      7293 &      0.162 &      2.176 &      324.3 $\pm$   19.4 &        0.4 $\pm$   0.15 &      -4.38 $\pm$   2.98 &       0.87 $\pm$   0.05 \\
      7293 &      2.176 &      3.008 &      254.2 $\pm$   15.8 &       1.02 $\pm$   0.29 &      -2.75 $\pm$   0.34 &       1.75 $\pm$   0.11 \\
      7293 &      3.008 &      3.776 &      224.7 $\pm$     15 &       0.89 $\pm$   0.26 &      -2.22 $\pm$   0.42 &       2.03 $\pm$   0.14 \\
      7293 &      3.776 &      5.568 &      175.6 $\pm$   8.51 &       1.05 $\pm$   0.24 &      -2.72 $\pm$    0.2 &       0.94 $\pm$   0.05 \\
      7293 &      5.568 &      7.552 &      167.5 $\pm$   6.31 &       0.77 $\pm$   0.19 &      -3.94 $\pm$   0.94 &       0.57 $\pm$   0.03 \\
      7293 &      7.552 &     11.008 &      139.9 $\pm$   4.65 &       0.99 $\pm$   0.21 &      -3.45 $\pm$   0.39 &       0.44 $\pm$   0.02 \\
      7293 &     11.008 &      13.76 &      119.2 $\pm$   5.43 &        1.1 $\pm$   0.35 &      -3.24 $\pm$   0.36 &       0.32 $\pm$   0.02 \\
      7293 &      13.76 &     18.688 &      90.75 $\pm$   5.14 &       1.76 $\pm$   0.64 &      -2.74 $\pm$   0.16 &       0.24 $\pm$   0.01 \\
     7475 &      3.072 &       4.16 &      291.5 $\pm$   88.8 &      -1.42 $\pm$   0.11 &      -2.61 $\pm$   1.41 &        1.2 $\pm$   0.37 \\
      7475 &       4.16 &      4.992 &        200 $\pm$   35.8 &      -1.13 $\pm$   0.14 &      -2.31 $\pm$   0.31 &       1.64 $\pm$   0.29 \\
      7475 &      4.992 &      5.696 &        212 $\pm$   41.3 &      -1.19 $\pm$   0.13 &      -2.24 $\pm$   0.27 &       2.12 $\pm$   0.41 \\
      7475 &      5.696 &        6.4 &      187.8 $\pm$   28.8 &      -1.14 $\pm$   0.13 &      -2.13 $\pm$   0.26 &       2.49 $\pm$   0.38 \\
      7475 &        6.4 &       7.04 &      207.5 $\pm$   45.3 &      -1.34 $\pm$   0.12 &      -2.15 $\pm$   0.28 &       2.58 $\pm$   0.56 \\
      7475 &       7.04 &      7.744 &        177 $\pm$   22.3 &      -1.24 $\pm$   0.11 &      -2.27 $\pm$   0.56 &       2.24 $\pm$   0.28 \\
      7475 &      7.744 &      8.448 &      142.3 $\pm$   22.8 &      -1.22 $\pm$   0.16 &      -2.24 $\pm$   0.17 &       2.19 $\pm$   0.35 \\
      7475 &      8.448 &      9.088 &      186.3 $\pm$   37.2 &      -1.27 $\pm$   0.14 &       -2.2 $\pm$   0.21 &       2.44 $\pm$   0.49 \\
      7475 &      9.088 &      9.792 &      178.7 $\pm$   17.7 &      -1.33 $\pm$   0.09 &      -6.48 $\pm$     72 &        1.6 $\pm$   0.16 \\
      7475 &      9.792 &     10.496 &      139.5 $\pm$   27.6 &      -1.25 $\pm$   0.19 &      -2.14 $\pm$   0.13 &       2.36 $\pm$   0.47 \\
      7475 &     10.496 &       11.2 &      162.7 $\pm$   19.9 &      -1.28 $\pm$   0.12 &      -2.78 $\pm$   0.72 &       1.62 $\pm$    0.2 \\
      7475 &       11.2 &     11.968 &      128.5 $\pm$   18.3 &      -1.22 $\pm$   0.17 &       -2.3 $\pm$   0.17 &       1.91 $\pm$   0.27 \\
      7475 &     11.968 &     12.672 &        139 $\pm$   11.6 &      -1.24 $\pm$   0.12 &      -3.51 $\pm$   2.37 &       1.38 $\pm$   0.12 \\
      7475 &     12.672 &      13.44 &      140.1 $\pm$   19.5 &      -1.12 $\pm$   0.17 &      -2.35 $\pm$   0.22 &       1.66 $\pm$   0.23 \\
      7475 &      13.44 &     14.272 &      128.2 $\pm$    9.6 &      -1.37 $\pm$   0.11 &      -8.56 $\pm$      5 &       1.24 $\pm$   0.09 \\
      7475 &     14.272 &     15.168 &      134.2 $\pm$   16.3 &      -1.61 $\pm$    0.1 &     -16.72 $\pm$      8 &       1.38 $\pm$   0.17 \\
      7475 &     15.168 &     16.064 &      99.57 $\pm$   10.4 &       -0.8 $\pm$   0.28 &       -2.4 $\pm$   0.15 &       1.21 $\pm$   0.18 \\
      7475 &     16.064 &     17.088 &      128.4 $\pm$   17.6 &      -1.64 $\pm$   0.11 &     -11.77 $\pm$      8 &       1.19 $\pm$   0.16 \\
      7475 &     17.088 &     18.176 &      107.6 $\pm$   10.3 &      -1.56 $\pm$   0.12 &        -23 $\pm$      8 &       0.99 $\pm$   0.09 \\
      7475 &     18.176 &     19.328 &      115.7 $\pm$   20.1 &      -1.66 $\pm$   0.14 &       -2.7 $\pm$   1.16 &       1.03 $\pm$   0.18 \\
      7475 &     19.328 &      22.08 &      88.74 $\pm$   6.17 &      -1.36 $\pm$   0.15 &      -2.78 $\pm$    0.4 &       0.66 $\pm$   0.05 \\
      7475 &      22.08 &      25.28 &      84.39 $\pm$   4.69 &      -1.32 $\pm$   0.13 &      -9.93 $\pm$      8 &       0.46 $\pm$   0.03 \\
      7475 &      25.28 &     29.248 &      73.83 $\pm$   6.05 &      -1.39 $\pm$   0.17 &     -19.39 $\pm$      8 &       0.32 $\pm$   0.03 \\
      7588 &      0.029 &      1.216 &      187.4 $\pm$   25.7 &      -0.35 $\pm$   0.26 &      -2.67 $\pm$   0.71 &       0.62 $\pm$   0.15 \\
      7588 &      1.216 &      2.176 &      91.72 $\pm$   8.32 &       1.25 $\pm$    0.7 &      -2.31 $\pm$   0.12 &       0.88 $\pm$   0.13 \\
      7588 &      2.176 &      3.136 &      78.56 $\pm$   6.22 &       0.53 $\pm$   0.58 &       -2.2 $\pm$   0.13 &       1.12 $\pm$   0.14 \\
      7588 &      3.136 &      4.224 &      67.31 $\pm$   4.19 &       0.82 $\pm$   0.68 &      -2.71 $\pm$   0.13 &       0.63 $\pm$   0.07 \\
      7588 &      4.224 &       5.44 &      62.78 $\pm$   3.72 &       1.47 $\pm$   0.89 &      -2.82 $\pm$   0.14 &       0.48 $\pm$   0.05 \\
      7588 &       5.44 &       6.72 &      56.13 $\pm$   4.62 &       1.02 $\pm$   1.16 &      -2.61 $\pm$   0.12 &        0.5 $\pm$   0.07 \\
      7588 &       6.72 &      8.384 &      54.95 $\pm$   3.43 &       1.47 $\pm$   1.23 &      -2.98 $\pm$   0.18 &       0.29 $\pm$   0.03 \\
      7711 &      0.032 &      1.088 &        485 $\pm$   74.7 &       -0.4 $\pm$   0.13 &      -2.68 $\pm$   0.46 &       1.85 $\pm$   0.37 \\
      7711 &      1.088 &      1.856 &      261.9 $\pm$   25.1 &       0.06 $\pm$   0.18 &      -2.21 $\pm$   0.23 &       2.35 $\pm$   0.32 \\
      7711 &      1.856 &      2.624 &      213.4 $\pm$   18.1 &      -0.15 $\pm$   0.17 &      -2.04 $\pm$   0.29 &       2.62 $\pm$   0.34 \\
      7711 &      2.624 &      3.456 &        174 $\pm$   16.5 &       -0.4 $\pm$   0.18 &      -2.42 $\pm$   0.22 &       1.48 $\pm$   0.21 \\
      7711 &      3.456 &       4.48 &      128.2 $\pm$   9.26 &      -0.38 $\pm$   0.21 &      -2.79 $\pm$   0.32 &       0.82 $\pm$    0.1 \\
      7711 &       4.48 &      5.824 &      100.9 $\pm$   7.57 &      -0.49 $\pm$   0.28 &      -2.75 $\pm$   0.27 &       0.58 $\pm$   0.07 \\
      7711 &      5.824 &      7.808 &      64.81 $\pm$   5.24 &       0.01 $\pm$   0.73 &      -2.71 $\pm$   0.18 &       0.33 $\pm$   0.05 \\
      7711 &      7.808 &      13.76 &      48.74 $\pm$   3.23 &       0.68 $\pm$   1.42 &      -3.02 $\pm$   0.21 &       0.13 $\pm$   0.02 \\
      7711 &      13.76 &     28.672 &      31.95 $\pm$   18.3 &       0.16 $\pm$   11.7 &      -2.82 $\pm$   0.31 &       0.04 $\pm$   0.06 \\
\enddata
\end{deluxetable}
\end{center}

\begin{thebibliography}{}
\bibitem[Amati et al.(2009)]{2009A&A...508..173A} Amati, L., Frontera, F., \& Guidorzi, C.\ 2009, \aap, 508, 173
\bibitem[Amati et al.(2002)]{2002A&A...390...81A} Amati, L., et al.\ 2002, \aap, 390, 81
\bibitem[Amati(2006)]{2006MNRAS.372..233A} Amati, L.\ 2006, \mnras, 372, 233
\bibitem[Band et al.(1993)]{1993ApJ...413..281B} Band, D., et al.\ 1993, \apj, 413, 281
\bibitem[Band \& Preece(2005)]{2005ApJ...627..319B} Band, D.~L., \& Preece, R.~D.\ 2005, \apj, 627, 319
\bibitem[Bhat et al.(1994)]{1994ApJ...426..604B} Bhat, P.~N., Fishman, G.~J., Meegan, C.~A., Wilson, R.~B., Kouveliotou, C., Paciesas, W.~S.,
Pendleton, G.~N., \& Schaefer, B.~E.\ 1994, \apj, 426, 604
\bibitem[Borgonovo \& Ryde(2001)]{2001ApJ...548..770B} Borgonovo, L., \& Ryde, F.\ 2001, \apj, 548, 770
\bibitem[Crider et al.(1999)]{1999A&AS..138..401C} Crider, A., Liang, E.~P., Preece, R.~D., Briggs, M.~S., Pendleton, G.~N., Paciesas, W.~S., Band, D.~L., \& Matteson, J.~L.\ 1999, \aaps, 138, 401
\bibitem[Dermer(2004)]{2004ApJ...614..284D} Dermer, C.~D.\ 2004, \apj, 614, 284
\bibitem[Dong et al. (2010)]{Dong2010}Dong, W., Liang, E. W., Lu, R. J.,
Science China G, 2010, 53 (S1), 78
\bibitem[Fenimore et al.(1995)]{1995ApJ...448L.101F} Fenimore, E.~E., in 't Zand, J.~J.~M., Norris, J.~P., Bonnell, J.~T.,
\& Nemiroff, R.~J.\ 1995, \apjl, 448, L101
\bibitem[Firmani et al.(2009)]{2009MNRAS.393.1209F} Firmani, C., Cabrera, J.~I., Avila-Reese, V., Ghisellini, G., Ghirlanda, G., Nava, L.,
\& Bosnjak, Z.\ 2009, \mnras, 393, 1209
\bibitem[Ford et al.(1995)]{1995ApJ...439..307F} Ford, L.~A., et al.\ 1995, \apj, 439, 307
\bibitem[Ghirlanda et al.(2005)]{2005MNRAS.360L..45G} Ghirlanda, G., Ghisellini, G., Firmani, C., Celotti, A.,
\& Bosnjak, Z.\ 2005, \mnras, 360, L45
\bibitem[Ghirlanda et al.(2010)]{2010A&A...511A..43G} Ghirlanda, G., Nava, L., \& Ghisellini, G.\ 2010, \aap, 511, A43
\bibitem[Ghirlanda et al.(2005)]{2005MNRAS.361L..10G} Ghirlanda, G., Ghisellini, G., \& Firmani, C.\ 2005, \mnras, 361, L10
\bibitem[Ghirlanda et al.(2004)]{2004ApJ...616..331G} Ghirlanda, G., Ghisellini, G., \& Lazzati, D.\ 2004, \apj, 616, 331
\bibitem[Golenetskii et al.(1983)]{1983Natur.306..451G} Golenetskii, S.~V., Mazets, E.~P., Aptekar, R.~L., \& Ilinskii, V.~N.\ 1983, \nat, 306, 451
\bibitem[Kaneko et al.(2006)]{2006ApJS..166..298K} Kaneko, Y., Preece, R.~D., Briggs, M.~S., Paciesas, W.~S., Meegan, C.~A.,
\& Band, D.~L.\ 2006, \apjs, 166, 298
\bibitem[Kargatis et al.(1994)]{1994ApJ...422..260K} Kargatis, V.~E., Liang, E.~P., Hurley, K.~C., Barat, C., Eveno, E.,
\& Niel, M.\ 1994, \apj, 422, 260
\bibitem[Kobayashi et al.(1997)]{1997ApJ...490...92K} Kobayashi, S., Piran, T., \& Sari, R.\ 1997, \apj, 490, 92
\bibitem[Lei et al.(2007)]{2007A&A...468..563L} Lei, W.~H., Wang, D.~X., Gong, B.~P., \& Huang, C.~Y.\ 2007, \aap, 468, 563
\bibitem[Liang et al.(2004)]{2004ApJ...606L..29L} Liang, E.~W., Dai, Z.~G., \& Wu, X.~F.\ 2004, \apjl, 606, L29
\bibitem[Liang \& Kargatis(1996)]{1996Natur.381...49L} Liang, E., \& Kargatis, V.\ 1996, \nat, 381, 49
\bibitem[Liang et al.(2002)]{2002PASJ...54....1L} Liang, E.-W., Xie, G.-Z.,\& Su, C.-Y.\ 2002, \pasj, 54, 1
\bibitem[Liang \& Zhang(2005)]{2005ApJ...633..611L} Liang, E. W., \& Zhang, B.\ 2005, \apj, 633, 611
\bibitem[Liang et al.(2009)]{2009arXiv0912.4800L} Liang, E.-W., Yi, S.-X., Zhang, J., LV, H.-J., Zhang, B.-B., \& Zhang, B.\ 2009, \apj, submitted
(arXiv:0912.4800).
\bibitem[Liu et al.(2010)]{2010arXiv1003.4883L} Liu, T., Liang, E.-W., Gu, W.-M., Zhao, X.-H., Dai, Z.-G., \& Lu, J.-F.\ 2010, arXiv:1003.4883
\bibitem[Lu \& Liang(2010)]{2010ScChG..53..163L} Lu, R., \& Liang, E.\ 2010, Science in China G: Physics and Astronomy, 53, 163
\bibitem[Medvedev(2006)]{2006ApJ...637..869M} Medvedev, M.~V.\ 2006, \apj, 637, 869
\bibitem[Mallozzi(2005)]{2005}Mallozzi, R. S., Preece, R. D., \& Briggs, M. S. 2005, RMFIT, A
Lightcurve and Spectral Analysis Tool, ( Huntsville: Univ. Alabama)
\bibitem[Nakar \& Piran(2005)]{2005MNRAS.360L..73N} Nakar, E., \& Piran, T.\ 2005, \mnras, 360, L73
\bibitem[Norris et al.(1986)]{1986ApJ...301..213N} Norris, J.~P., Share,
G.~H., Messina, D.~C., Dennis, B.~R., Desai, U.~D., Cline, T.~L., Matz, S.~M.,
\& Chupp, E.~L.\ 1986, \apj, 301, 213
\bibitem[Ohno et al.(2009)]{2009PASJ...61..201O} Ohno, M., Ioka, K.,Yamaoka, K., Tashiro, M., Fukazawa, Y.,
\& Nakagawa, Y.~E.\ 2009, \pasj, 61, 201
\bibitem[Panaitescu et al.(1998)]{1998ApJ...503..314P} Panaitescu, A.,
Meszaros, P., \& Rees, M.~J.\ 1998, \apj, 503, 314
\bibitem[Panaitescu\& Vestrand(2008)]{2008MNRAS.387..497P} Panaitescu, A., \& Vestrand, W.~T.\ 2008, \mnras, 387, 497
\bibitem[Peng et al.(2009)]{2009ApJ...698..417P} Peng, Z.~Y., Ma, L., Zhao, X.~H., Yin, Y., Fang, L.~M., \& Bao, Y.~Y.\ 2009, \apj, 698, 417
\bibitem[Portegies Zwart et al.(1999)]{1999A&AS..138..503P} Portegies Zwart, S.~F., Lee, C.-H., \& Lee, H.~K.\ 1999, \aaps, 138, 503
\bibitem[Preece et al.(1998)]{1998ApJ...506L..23P} Preece, R.~D., Briggs,
M.~S., Mallozzi, R.~S., Pendleton, G.~N., Paciesas, W.~S., \& Band, D.~L.\
1998, \apjl, 506, L23
\bibitem[Qin(2008)]{2008ApJ...683..900Q} Qin, Y.-P.\ 2008, \apj, 683, 900
\bibitem[Qin(2002)]{2002A&A...396..705Q} Qin, Y.-P.\ 2002, \aap, 396, 705
\bibitem[Reynoso et al.(2006)]{2006A&A...454...11R} Reynoso, M.~M., Romero, G.~E., \& Sampayo, O.~A.\ 2006, \aap, 454, 11
\bibitem[Sakamoto et al.(2006)]{2006ApJ...636L..73S} Sakamoto, T., et al.\ 2006, \apjl, 636, L73
\bibitem[Shahmoradi \& Nemiroff(2009)]{2009arXiv0904.1464S} Shahmoradi, A., \& Nemiroff, R.~J.\ 2009, arXiv:0904.1464
\bibitem[Shen et al.(2005)]{2005MNRAS.362...59S} Shen, R.-F., Song, L.-M.,
\& Li, Z.\ 2005, \mnras, 362, 59
\bibitem[Toma et al.(2007)]{2007ApJ...659.1420T} Toma, K., Ioka, K.,
Sakamoto, T., \& Nakamura, T.\ 2007, \apj, 659, 1420
\bibitem[Wei \& Gao(2003)]{2003MNRAS.345..743W} Wei, D.~M., \& Gao, W.~H.\ 2003, \mnras, 345, 743
\bibitem[Yonetoku et al.(2004)]{2004ApJ...609..935Y} Yonetoku, D.,
Murakami, T., Nakamura, T., Yamazaki, R., Inoue, A.~K., \& Ioka, K.\ 2004,
\apj, 609, 935
\bibitem[Zhang et al.(2009)]{2009ApJ...690L..10Z} Zhang, B.-B., Zhang, B.,
Liang, E.-W., \& Wang, X.-Y.\ 2009, \apjl, 690, L10
\bibitem[Zhang \& M{\'e}sz{\'a}ros(2002)]{2002ApJ...581.1236Z} Zhang, B., \& M{\'e}sz{\'a}ros,
P.\ 2002, \apj, 581, 1236
\bibitem[Zhang\& M{\'e}sz{\'a}ros(2004)]{2004IJMPA..19.2385Z} Zhang, B., \& M{\'e}sz{\'a}ros,
P.\ 2004, International Journal of Modern Physics A, 19, 2385
\end{thebibliography}
\end{document}